\documentclass{amsart}

\newtheorem{theorem}{Theorem}[section]

\theoremstyle{definition}
\newtheorem{definition}[theorem]{Definition}
\newtheorem{example}[theorem]{Example}

\theoremstyle{remark}

\usepackage{epsfig}

\reversemarginpar

\begin{document}

\newtheorem{conjecture}{Conjecture}[section]
\newtheorem{corollary}{Corollary}[section]
\newtheorem{Note}{Note}[section]

\theoremstyle{definition}
\newtheorem{rules}{Rule}[section]

\theoremstyle{remark}


\title[Evaluation of Integrals by Brackets]{Definite 
integrals by the method of brackets. Part 1}

\author{Ivan Gonzalez}
\address{Departmento de Fisica, Pontificia 
Universidad Catolica de Santiago, Chile}
\email{ivan.gonzalez@usm.cl}
\email{igonzalez@fis.puc.cl}

\author{Victor H. Moll}
\address{Department of Mathematics,
Tulane University, New Orleans, LA 70118}
\email{vhm@math.tulane.edu}

\thanks{The first author was partially funded by Fondecyt (Chile), Grant 
number $3080029$. The work of the second author was partially funded by
$\text{NSF-DMS } 0409968$.}



\numberwithin{equation}{section}

\newcommand{\imagpart}{\mathop{\rm Im}\nolimits}
\newcommand{\realpart}{\mathop{\rm Re}\nolimits}
\newcommand{\no}{\noindent}
\newcommand{\ift}{\int_{0}^{\infty}}
\newcommand{\ione}{\int_{0}^{1}}
\newcommand{\eqf}{\stackrel{\bullet}{=}}

\begin{abstract}
A new heuristic 
method for the evaluation of definite integrals is presented. This 
{\em method of brackets} has its origin in methods developed for the 
evaluation of Feynman diagrams. We describe the operational rules and 
illustrate the method with several examples.  The method of brackets reduces 
the evaluation of a large class of definite integrals to the solution of 
a linear system of equations. 
\end{abstract}

\maketitle

\section{Introduction} \label{sec-intro}
\setcounter{equation}{0}

The problem of analytic evaluations of definite integrals has been of 
interest to scientists since Integral Calculus was developed. The central 
question can be stated vaguely as follows: \\

\begin{center}
{\em given a class of 
functions} $\mathfrak{F}$ {\em and an 
interval} $[a,b] \subset \mathbb{R}$, {\em express the integral of} $f \in 
\mathfrak{F}$ 
\begin{equation}
I = \int_{a}^{b} f(x) \, dx,
\nonumber
\end{equation}
\noindent
{\em in terms of the special values of functions in an enlarged class}
$\mathfrak{G}$. 
\end{center}

\medskip

For instance, by elementary 
arguments it is possible to show that if $\mathfrak{F}$ is the 
class of rational functions, then the enlarged class $\mathfrak{G}$ can be 
obtained by including logarithms and inverse trigonometric 
functions. G. Cherry has 
discussed in \cite{cherry1}, \cite{cherry2} and \cite{cherry5}
extensions of this classical paradigm.
The following results illustrate the idea: 
\begin{equation}
\int \frac{x^{3} \, dx}{\ln(x^{2}-1)} = 
\frac{1}{2} \text{li}(x^{4}-2x^{2}+1) + \frac{1}{2} \text{li}(x^{2}-1),
\label{ex-1}
\end{equation}
\noindent
but
\begin{equation}
\int \frac{x^{2} \, dx}{\ln(x^{2}-1)} 
\end{equation}
\noindent
can not be written in terms of elementary functions and the 
logarithmic integral
\begin{equation}
\text{li}(x) := \int \frac{x \, dx}{\ln x}
\end{equation}
\noindent
that appears in (\ref{ex-1}). The reader will find in \cite{bronstein2} 
the complete theory behind integration in terms of elementary functions. 

Methods for the evaluation of definite integrals were also developed since
the early stages of Integral Calculus. Unfortunately, these are mostly
ad-hoc procedures and a general theory needs to be developed. The method 
proposed in this paper represents a new addition to these procedures. 

The evaluations of definite integrals have been collected in 
tables of integrals. 
The earliest volume available to the authors 
is \cite{bierens1}, compiled by Bierens de Haan who also presented 
in \cite{bierens2} a survey of the methods employed in the verification of 
the entries. These tables form the main source for the 
popular volume by I. S. Gradshteyn and I. M. Ryzhik \cite{gr}. 
Naturally any document containing a 
large number of entries, such as the table \cite{gr} or the encyclopedic
treatise \cite{prudnikov1}, is 
likely to contain errors. 
For instance, the appealing 
integral
\begin{equation}
I = \ift \frac{dx}{(1+x^2)^{3/2} \left[ \varphi(x) + \sqrt{\varphi(x) } 
\right]^{1/2}} = \frac{\pi}{2 \sqrt{6}},
\label{wrong}
\end{equation}
\noindent
with 
\begin{equation}
\varphi(x) = 1 + \frac{4x^{2}}{3(1+x^{2})^{2}},
\end{equation}
that appears as entry $3.248.5$ in \cite{gr6}, the sixth 
edition of the mentioned table, is incorrect. The numerical value of $I$ is 
$0.666377$ and the right hand side of (\ref{wrong}) is about $0.641275$. 
The table \cite{gr} is in the process of being revised. After we informed 
the editors of the error in $3.248.5$, it was taken out. There is no entry
$3.248.5$ in  \cite{gr}. 
At the present time, we are unable to evaluate the integral $I$. 

The revision of integral tables 
is nothing new. C. F. Lindman \cite{lindman1} compiled a long list of 
errors from the table by Bierens de Haan \cite{bierens3}. The editors of 
\cite{gr} maintain the webpage 
\begin{verbatim}                     http://www.mathtable.com/gr/\end{verbatim}
\noindent
where the corrections to the table are 
stored. The second author 
has began in \cite{moll-gr5,moll-gr7,moll-gr1,moll-gr2,moll-gr3,
moll-gr4,moll-gr6,moll-gr8} a systematic verification of the entries in 
\cite{gr}. It is in this task that the method proposed in the present 
article becomes a valuable tool. 

The {\em method of brackets} presented here, even 
though it is heuristic and 
still lacking a rigorous description, is quite powerful. Moreover, it
is quite simple to work with: the evaluation of a definite integral is 
reduced to solving a linear system of equations. Many of the entries
of \cite{gr} can be derived using this method. 
The basic idea behind it is the assignement
of a {\em bracket} $\langle{a \rangle}$ to any parameter $a$. 
This is a symbol associated to the divergent integral 
\begin{equation}
\ift x^{a -1} \, dx.
\end{equation}
\noindent
The formal rules for operating with these brackets are described in Section 
\ref{sec-bracket} and their justification is work-in-progress, we expect 
to report in the near future.  The rest of the paper 
provides a list of examples 
illustrating  the new technique. 

Given a formal sum 
\begin{equation}
f(x) = \sum_{n=0}^{\infty} a_{n}x^{\alpha n + \beta -1}
\end{equation}
\noindent
we associate to the integral of $f$ a {\em bracket series} written as 
\begin{equation}
\ift f(x) \, dx \eqf \sum_{n} a_{n} \langle{ \alpha n + \beta \rangle}, 
\end{equation}
\noindent 
to keep in mind the formality of the method described in this paper. 
Convergence 
issues are ignored at the present time. Moreover only integrals over the 
half-line $[0, \infty)$ will be considered.  \\

\noindent
{\bf Note}. In the evaluation of these formal sums, the index 
$n \in \mathbb{N}$ will be replaced by a number $n^{*}$ 
defined by the vanishing 
of the bracket. Observe that it is possible that $n^{*} \in \mathbb{C}$. 
For book-keeping purposes, specially in cases
with many indices, we write $\displaystyle{\sum_{n}}$ instead of the usual
$\displaystyle{\sum_{n=0}^{\infty}}$. After the brackets are eliminated, those 
indices that remain recover their original nature.  \\

The rules of 
operation described below assigns a {\em value}
 to the bracket series. The claim is 
that for a large class of integrands, including all the examples 
described here, this formal procedure 
provides the actual value of the integral.  Many
of the examples involve the hypergeometric function 
\begin{equation}
{_{p}F_{q}}(z) := \sum_{n=0}^{\infty} 
\frac{(a_{1})_{n} \, (a_{2})_{n} \cdots (a_{p})_{n}}
{(b_{1})_{n} \, (b_{2})_{n} \cdots (b_{q})_{n}} \frac{z^{n}}{n!}.
\end{equation}
\noindent
This series converges absolutely for all $z \in \mathbb{C}$ if $p \leq q$ and 
for $|z|< 1$ if $p = q+1$. The series diverges for all $z \neq 0$ if 
$p > q+1$ unless the series terminates.  The special case $p = q+1$ is of 
great interest. In this special case and with $|z|=1$, the series 
\begin{equation}
{_{q+1}F_{q}}(a_{1}, \cdots, a_{q+1}; b_{1}, \cdots, b_{q};z)
\end{equation}
\noindent
converges absolutely if $\realpart{ \left(\sum b_{j} - \sum a_{j}
\right) } >0$. The series converges conditionally if $z = e^{i \theta} \neq 1$
and $0 \geq \realpart{ \left(\sum b_{j} - \sum a_{j} \right)} > -1$ 
and the series diverges if 
$\realpart{ \left(\sum b_{j} - \sum a_{j} \right)} \leq  -1$. 

{{
\begin{figure}[ht]
\begin{center}
\centerline{\epsfig{file=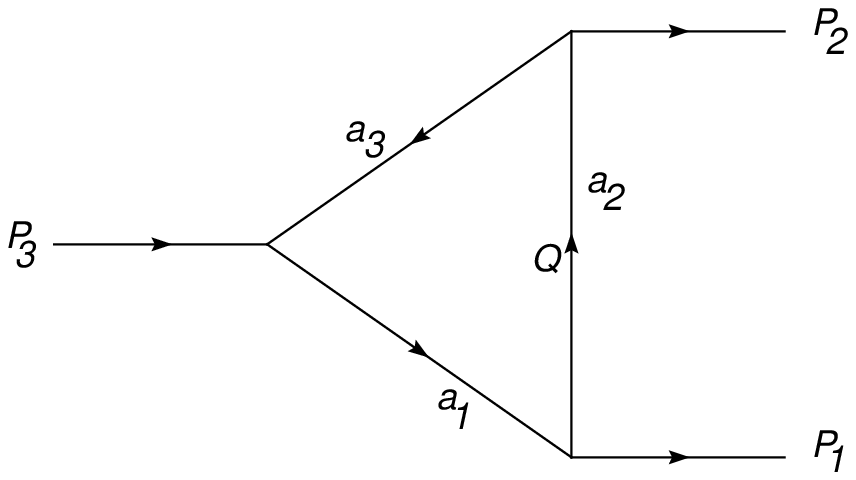,width=20em,angle=0}}
\caption{The triangle}
\label{figure0}
\end{center}
\end{figure}
}}

The last section of this paper 
employs the method of brackets to evaluate
certain definite integrals associated to a Feynman diagram. From the 
present point of view, a {\em Feynman diagram} is simply a generic graph $G$
that contains $E+1$ {\em external} lines and $N$ {\em internal} lines or 
{\em propagators} and $L$ {\em loops}. All but one of these external lines 
are assumed independent. The internal and 
external lines represent particles that transfer momentum among the vertices 
of the diagram. Each of these particles 
carries a {\em mass} $m_{i} \geq 0$ for $i=1,\cdots, N$. The 
vertices represent the interaction of these
particles and conservation of momentum at each vertex assigns the momentum 
corresponding to the internal lines.  A Feynman diagram has an associated 
integral given by the {\em parametrization} of the 
diagram. For example, in Figure \ref{figure0}
we have three external lines represented by the momentum $P_{1}, \, P_{2}, 
\, P_{3}$ and one loop. The parameters $a_{i}$ are arbitrary real 
numbers. The integral 
associated to this diagram is given by
\begin{eqnarray}
G  & = & \frac{(-1)^{-D/2}}{\Gamma(a_{1})\Gamma(a_{2})\Gamma(a_{3})} 
\ift \ift \ift \frac{x_{1}^{a_{1}-1} x_{2}^{a_{2}-1} x_{3}^{a_{3}-1} }
{(x_{1}+x_{2}+x_{3})^{D/2}} 
\nonumber \\
& \times  & \text{exp}(x_{1}m_{1}^{2} + x_{2}m_{2}^{2} + x_{3}m_{3}^{2}) 
\, \, \text{exp} \left( - 
\frac{C_{11}P_{1}^{2} + 2C_{12}P_{1} \cdot P_{2} + C_{22}P_{2}^{2} }
{x_{1}+x_{2}+x_{3}} \right) {\mathbf{dx}},
\nonumber
\end{eqnarray}
\noindent
where ${\mathbf{dx}} := dx_{1} \, dx_{2} \, dx_{3}$. The 
evaluation of this integral in terms of the variables $P_{i} \in 
{\mathbb{R}}^{4}, \, m_{i} \in \mathbb{R}$ and $a_{i} \in \mathbb{R}$
is the {\em solution of the Feynman diagram}. The functions $C_{ij}$ are 
polynomials described in Section \ref{sec-feynman}. 

The method of brackets presented here has its origin in quantum field
theory (QFT). A 
version of the method of brackets was developed to address one of 
the fundamental questions in QFT: the evaluation of loop integrals arising 
from Feynman diagrams. As described above, these 
are directed graphs depicting the interaction
of particles in the model. The loop integrals depend on the dimension $D$
and one of the (many) intrinsic difficulties 
is related to their divergence at $D=4$, the dimension of 
the physical world. A correction to this problem is obtained by 
taking $D= 4 - 2 \epsilon$ and considering a Laurent expansion in powers of 
$\epsilon$. This is called the {\em dimensional regularization} \cite{bollini}
and the parameter $\epsilon$ is the {\em dimensional regulator}. 

The method of brackets discussed in this paper is based on previous results by
I. G. Halliday,  R. M. Ricotta and G. V. Dunne 
\cite{dunne-1987}, 
\cite{dunne-1989} and 
\cite{halliday-1987}. The work involves an analytic extension of  $D$ to 
negative values, so the method was labelled NDIM (negative dimensional 
integration method). The validity of this continuation is based on the 
observation that the objects associated to a Feynman diagram (loop integrals 
as well as the functions linked to propagators) are analytic in the dimension
$D$.  A. Suzuki and A. Schmidt employed this technique to the evaluation of
diagrams with two loops \cite{suzuki-twoa}, \cite{suzuki-twoa1}; three loops
\cite{suzuki-three}; tensorial integrals \cite{suzuki-tensor} and massive 
integrals with one loop \cite{suzuki-massive2}, \cite{suzuki-massive}, 
\cite{suzuki-massive1}. An extensive use of this 
method as well as an analysis of the solutions 
was provided by C. Anastasiou and E. Glover in \cite{anastasiou-a} and 
\cite{anastasiou-b}. The conclusion of these studies is that the NDIM method
is inadequate to the evaluation of Feynman diagrams with an arbitrary number
of loops. The proposed solutions involve hypergeometric functions with a 
large number of parameters. By establishing new procedural rules 
I. Gonzalez and I.Schmidt \cite{gonzalez-2007} and \cite{gonzalez-2008} 
have concluded that the modification of the previous procedures permits now
the evaluation of more complex Feynman diagrams. One of the results of 
\cite{gonzalez-2007}, \cite{gonzalez-2008} is the justification of the 
method of brackets in terms of arguments derived from fractional calculus. 
The authors have given NDIM the alternative name IBFE (Integration by 
Fractional Expansion). 

From the mathematical 
point of view, the NDIM method has been used to provide evaluation of a
very limited type of integrals \cite{suzuki-int2}, \cite{suzuki-int1}. The 
examples presented in this paper show great flexibility of the method 
of brackets. A systematic study of integrals arising from Feynman diagrams 
is in preparation. 

\section{A detour on definite integrals} \label{sec-detour}
\setcounter{equation}{0}

The literature contains a large variety of techniques for 
the evaluation of definite integrals. 
Elementary techniques are surveyed in classical texts such as 
\cite{edwards2} and \cite{fichtenholz1}. The text
\cite{antimirov1} contains an excellent collection of problems solved by 
the method of contour integration. The reader will find in 
\cite{irrbook} a discussion of several elementary analytic methods involved
in the evaluation of integrals. 

It is hard to predict the 
type of techniques required for the evaluation of a specific definite 
integral. For instance,  
\cite{vardi1} contains a detailed account of the 
proof of
\begin{equation}
\int_{\pi/4}^{\pi/2} \ln \ln \tan x \, dx = \frac{\pi}{2} 
\ln \left( \frac{\sqrt{2 \pi} \Gamma \left( \tfrac{3}{4} \right) }
{\Gamma \left( \tfrac{1}{4} \right) } \right),
\end{equation}
\noindent
that appears as formula $4.229.7$ in \cite{gr}. 
This particular example involves the use of $L$-functions
\begin{equation}
L_{\chi}(s) = \sum_{n=1}^{\infty} \frac{\chi(n)}{n^{s}},
\end{equation}
\noindent
where $\chi$ is a character. This is a generalization of the Riemann zeta 
function $\zeta(s)$ (corresponding 
to $\chi \equiv 1$). Vardi's  technique 
has been extended in \cite{luis2} to provide a systematic study of 
integrals of the form 
\begin{equation}
I_{Q} = \ione Q(x) \ln \ln 1/x \, dx,
\end{equation}
\noindent
that gives evaluations such as 
\begin{eqnarray}
\ione \frac{x \ln \ln 1/x \, dx}{(x+1)^{2}} & = & 
\frac{1}{2} ( - \ln^{2}2 + \gamma - \ln \pi + \ln 2 ), \\
\ift \ln x \ln \tanh x \, dx  & = & 
\frac{\gamma \pi^{2}}{8} - \frac{3}{4} \zeta'(2) + \frac{\pi^{2} \ln 2}{12}.
\end{eqnarray}
\noindent
Here $\gamma = - \Gamma'(1)$ is Euler's constant.

A second class of examples appeared
during the evaluation of definite integrals
related to the Hurwitz zeta function 
\begin{equation}
\zeta(z,q) =  \sum_{n=0}^{\infty} \frac{1}{(n+q)^{z}}.
\end{equation}
\noindent
In \cite{espmoll1} the authors found an
evaluation that generalizes the classical integral
\begin{equation}
L_{1} := \ione \ln \Gamma(q) \, dq = \ln \sqrt{2 \pi},
\label{L1-value}
\end{equation}
\noindent
namely
\begin{equation}
L_{2} := \ione \ln^{2} \Gamma(q) \, dq = 
\frac{\gamma^{2}}{12} + \frac{\pi^{2}}{48} + \frac{\gamma L_{1}}{3} 
+ \frac{4}{3} L_{1}^{2} - \frac{A \zeta'(2)}{\pi^{2}} + 
\frac{\zeta''(2)}{2 \pi^{2}},
\end{equation}
\noindent
where $L_{1}$ is in (\ref{L1-value}) and 
$A = \gamma + \ln 2 \pi$. The natural 
next step, namely the evaluation of 
\begin{equation}
L_{3} := \ione \ln^{3} \Gamma(q) \, dq, 
\end{equation}
\noindent
remains to be completed. In \cite{espmoll3} and \cite{espmoll6} the reader 
will find a relation between $L_{3}$ and the Tornheim sums $T(m,k,n)$,  
for $m, k, n \in \mathbb{N}$. These sums are defined by
\begin{equation}
T(a,b,c) := \sum_{j_{1}=1}^{\infty} \sum_{j_{2}=1}^{\infty} 
\frac{1}{j_{1}^{a} \, j_{2}^{b} \, (j_{1} + j_{2} )^{c}}.
\end{equation}
\noindent
The special case 
\begin{equation}
T(n,0,m) =  \sum_{j_{2} > j_{1}}  \frac{1}{j_{1}^{m} j_{2}^{n}},
\end{equation}
\noindent
corresponds to the multiple zeta value (MZV) $\zeta(n,m)$ of 
depth $2$. The MZV is given by 
\begin{equation}
\zeta(n_{1},n_{2}, \cdots, n_{r}) := \sum_{j_{1} > j_{2} > \cdots > j_{r}} 
\frac{1}{j_{1}^{n_{1}} \, j_{2}^{n_{2}} \cdots j_{r}^{n_{r}} },
\end{equation}
\noindent
where the parameter $r$ is called the depth of the sum. These series were 
initially considered by Euler and have recently appeared in many different 
places. The reader will find in \cite{kreimer1} a description of how these
sums are connected to knots and Feynman diagrams. These diagrams
are a very rich source of interesting integrals. The last section of this 
paper is dedicated to the evaluation of some of these integrals by the 
method of brackets. 

The computation of hyperbolic volumes of $3$-manifolds provides a
different source of interesting integrals. Mostow's rigidity theorem states 
that a finite volume $3$-manifold has a unique hyperbolic structure. In 
particular its volume is a topological invariant. An interesting class of 
such $3$-manifolds is provided by 
hyperbolic knots or link complement in $S^{3}$. The reader will find 
information about this topic in the articles by C. Adams and J. Weeks in 
\cite{menasco}. It turns out that 
their hyperbolic structure can be given in 
terms of hyperbolic tetrahedra \cite{adams1}. Milnor \cite{milnor82} 
describes how the volume of these tetrahedra can be expressed in terms of 
the Clausen function 
\begin{equation}
\text{Cl}_{2}(\theta) := - \int_{0}^{\theta} \log | 2 \sin \frac{u}{2} | \, du. 
\label{clausen-1}
\end{equation}
\noindent
The reader will find in \cite{maclachlan-reid} a discussion on arithmetic 
properties of $3$-manifolds. In particular, Chapter 11 has up to date 
information on their volumes.

Zagier \cite{zagier2} provided an arithmetic version of these computations 
in his study of the Dedekind zeta function 
\begin{equation}
\zeta_{K}(s) := \sum_{\frak{a}} \frac{1}{N(\frak{a})^{s}},
\end{equation}
\noindent
for a number field 
$K$ that is not totally real. Here $N(\frak{a})$ is the norm of the 
ideal $\frak{a}$ and the sum runs over all the nonzero integral ideals of 
$K$. In the case of totally real number fields a classical result of 
Siegel shows that $\zeta_{K}(2m)$ is a rational multiple 
of $\pi^{2nm}/\sqrt{D}$, where $n$ and $D$ denote the degree and the 
discriminant of $K$, respectively. Little is known in the non-totally real 
situation. Zagier \cite{zagier2} 
proves that $\zeta_{K}(2)$ is given by a finite sum of values of 
\begin{equation}
A(x) := \int_{0}^{x} \frac{1}{1+t^{2}} \log \frac{4}{1+t^{2}} \, dt.
\end{equation}
\noindent
The function $A$ can be written as 
\begin{equation}
A(x) = \text{Cl}_{2}(\pi - 2 \tan^{-1}x) - \text{Cl}_{2}(\pi). 
\end{equation}
\noindent
Morever, he conjectured that $\zeta_{k}(2m)$ can be given in terms of 
\begin{equation}
A_{m}(x) := \frac{2^{2m-1}}{(2m-1)!} 
\ift \frac{t^{2m-1} \, dt}{x \sinh^{2}t + x^{-1} \cosh^{2}t}.
\end{equation}
\noindent
The conjecture is established in the special case where $K$ is an abelian 
extension of $\mathbb{Q}$. The example $K = \mathbb{Q}(\sqrt{-7})$ yields 
\begin{equation}
\zeta_{\mathbb{Q}(\sqrt{-7})}(2) = \frac{\pi^{2}}{3 \sqrt{7}} 
\left( 
A \left( \cot \tfrac{\pi}{7} \right) + 
A \left( \cot \tfrac{2 \pi}{7} \right) + 
A \left( \cot \tfrac{4 \pi}{7} \right)  \right)
\end{equation}
\noindent 
and also 
\begin{equation}
\zeta_{\mathbb{Q}(\sqrt{-7})}(2) =  \frac{2 \pi^{2}}{7 \sqrt{7}} 
\left( 2 A (\sqrt{7}) + 
A( \sqrt{7} + 2 \sqrt{3} ) + 
A(\sqrt{7} - 2 \sqrt{3}) \right), 
\end{equation}
\noindent
leading to the new Claussen identity
\begin{equation}
A \left( \cot \tfrac{\pi}{7} \right) + 
A \left( \cot \tfrac{2 \pi}{7} \right) + 
A \left( \cot \tfrac{4 \pi}{7} \right)  = 
\tfrac{6}{7} \left( 
2 A (\sqrt{7}) + 
A( \sqrt{7} + 2 \sqrt{3} ) + 
A(\sqrt{7} - 2 \sqrt{3}) \right). 
\nonumber 
\end{equation}
\noindent 
Zagier stated in $1986$ that there was no direct proof of this identity. To 
this day this has elluded considerable effort. The 
famous text of Lewin \cite{lewin1} has such parametric identities but it 
misses this one.
R. Crandall \cite{crandall-08}
has worked out a theory in which certain Claussen identities are seen to be 
equivalent to the vanishing of log-rational integrals. \\

J. Borwein and D. Broadhurst \cite{bor-broad1998} identified a large 
number of finite volume hyperbolic $3$-manifolds whose volumes are 
expressed in the form 
\begin{equation}
\frac{a}{b} \text{vol}(\frak{M}) = 
\frac{(-D)^{3/2} }{(2 \pi)^{2n-4}} \frac{\zeta_{K}(2)}{2 \zeta(2)}.
\label{bor-bro}
\end{equation}
\noindent 
Here $K$ is a field associated to the manifold $\frak{M}$ (the so-called 
invariant trace field) and $n$ and $D$ are 
the degree and discriminant of $K$, respectively.  The authors offer 
a systematic numerical study of the rational numbers 
$\displaystyle{\frac{a}{b}}$. The identity of Zagier described above yields 
the remarkable identity 
\begin{equation}
\int_{\pi/3}^{\pi/2} 
\ln \left| \frac{\tan t + \sqrt{7}}{\tan t - \sqrt{7}} \right| \, dt
= 
A( \sqrt{7}) + 
\frac{1}{2} A( \sqrt{7} + 2 \sqrt{3}) + 
\frac{1}{2} A( \sqrt{7} - 2 \sqrt{3}).
\nonumber
\end{equation}
\noindent 
This example corresponds to the link $6_{1}^{3}$ with discriminant $D= - 7$. 
Zagier's result gives $\displaystyle{\frac{a}{b}} = 2$ in (\ref{bor-bro}). \\

Coffey \cite{coffey2008}, \cite{coffey2009} 
 has studied the integral above, that 
also appears in the reduction of a 
multidimensional Feynman integral \cite{lunev1994}.  The goal is to produce 
a more direct proof of Zagier remarkable identity as well as the many 
others that have been numerically verified in \cite{bor-broad1998}. 

\medskip

The subject of evaluation of definite integrals has a rich history. We 
expect that the method of brackets developed in this paper will
expand the class of integrals that can be expressed in 
analytic  form. 

\section{The method of brackets} \label{sec-bracket}
\setcounter{equation}{0}

The method of brackets discussed in this paper is based on the assignment
of a {\em bracket} $\langle{a \rangle}$ the parameter $a$. In the
examples presented here $a \in \mathbb{R}$, but the extension to
$a \in \mathbb{C}$ is direct. The 
formal rules for operating with these brackets are described next.  

\begin{definition}
Let $f$ be a formal power series 
\begin{equation}
f(x) = \sum_{n=0}^{\infty} a_{n}x^{\alpha n+\beta-1}.
\label{series-f}
\end{equation}
\noindent
The symbol 
\begin{equation}
\ift f(x) \, dx \eqf \sum_{n} a_{n} \langle{ \alpha n + \beta \rangle}
\end{equation}
\noindent 
represents a {\em bracket series} assignement to the 
integral on the left. Rule \ref{rule-ass1} describes how to evaluate this
series. 
\end{definition}

\begin{definition}
The symbol
\begin{equation}
\phi_{n} := \frac{(-1)^{n}}{\Gamma(n+1)}
\end{equation}
\noindent 
will be called the {\em indicator of} $n$.
\end{definition}

\noindent
The symbol $\phi_{n}$ gives a simpler form for the bracket series associated 
to an integral. For example, 
\begin{equation}
\ift x^{a-1} e^{-x} \, dx \eqf \sum_{n} \phi_{n} \langle{ n + a \rangle}.
\end{equation}
\noindent
The integral is the  gamma function $\Gamma(a)$ and the right-hand side 
its bracket expansion. 

\begin{rules}
\label{rule-binom}
For $\alpha \in \mathbb{C}$, the expression 
\begin{equation}
(a_{1} + a_{2} + \cdots + a_{r})^{\alpha}
\end{equation}
\noindent
is assigned the bracket series 
\begin{equation}
\sum_{m_{1}, \cdots, m_{r}} \phi_{1,2,\cdots,r} \, a_{1}^{m_{1}} 
\cdots a_{r}^{m_{r}} 
\frac{ \langle{-\alpha + m_{1} + \cdots + m_{r} \rangle}}{\Gamma(-\alpha)},
\end{equation}
\noindent
where $\phi_{1,2,\cdots,r}$ is a short-hand notation for the product
$\phi_{m_{1}} \phi_{m_{2}} \cdots \phi_{m_{r}}$.
\end{rules}

\noindent
\begin{rules}
\label{rule-ass1}
The series of brackets
\begin{equation}
\sum_{n} \phi_{n} f(n) \langle{ a n + b \rangle} 
\end{equation}
\noindent
is given the {\em value} 
\begin{equation}
\frac{1}{a} f(n^{*}) \Gamma(-n^{*})
\end{equation}
\noindent
where $n^{*}$ solves the equation $an+b = 0$. 
\end{rules}

\noindent
\begin{rules}
A two-dimensional series of brackets
\begin{equation}
\sum_{n_{1}, n_{2}} \phi_{n_{1},n_{2}} f(n_{1},n_{2}) 
\langle{ a_{11} n_{1} + a_{12}n_{2}+c_{1} \rangle} 
\langle{ a_{21} n_{1} + a_{22}n_{2}+c_{2} \rangle} 
\end{equation}
\noindent
is assigned the {\em value} 
\begin{equation}
\frac{1}{|a_{11} a_{22} - a_{12}a_{21}|} f(n_{1}^{*}, n_{2}^{*}) \Gamma(-n_{1}^{*}) 
\Gamma( -n_{2}^{*})
\end{equation}
\noindent
where $(n_{1}^{*},n_{2}^{*})$ is the unique solution to the linear 
system 
\begin{eqnarray}
a_{11} n_{1} + a_{12}n_{2}+c_{1}  & = & 0,  \label{system-11} \\
a_{21} n_{1} + a_{22}n_{2}+c_{2} & = & 0, \nonumber
\end{eqnarray}
\noindent
obtained by the vanishing of the expressions in the brackets. A 
similar rule applies to higher dimensional series, that is,
\begin{equation}
\sum_{n_{1}} \cdots \sum_{n_{r}} \phi_{1,\cdots,r} 
f(n_{1},\cdots,n_{r}) 
\langle{ a_{11}n_{1}+ \cdots a_{1r}n_{r} + c_{1} \rangle} \cdots 
\langle{ a_{r1}n_{1}+ \cdots a_{rr}n_{r} + c_{r} \rangle} 
\nonumber
\end{equation}
\noindent
is assigned the value 
\begin{equation}
\frac{1}{| \text{det}(A) |} f(n_{1}^{*}, \cdots, n_{r}^{*}) 
\Gamma(-n_{1}^{*}) \cdots f(-n_{r}^{*}),
\end{equation}
\noindent
where $A$ is the matrix of coefficients 
$(a_{ij})$ and $\{ n_{i}^{*} \, \}$ is the solution of the 
linear system obtained by the vanishing of the brackets.  The value is not
defined if the matrix $A$ is not invertible.
\end{rules}

\begin{rules}
\label{rule-disc}
In the case where the assignment leaves free parameters, any 
divergent series in these parameters is discarded.  In case several choices
of free parameters are available, the series that converge in a common region 
are added to contribute to the integral. 
\end{rules}

A  typical place  to apply Rule \ref{rule-disc} is where  the 
hypergeometric functions ${_{p}F_{q}}$, with $p = q+1$, appear. In this
case the convergence of the series imposes restrictions on the 
internal parameters of the problem. 
Example \ref{example-bubble}, dealing 
with a Feynman diagram with a {\em bubble}, 
illustrates the latter part of this rule. \\

\noindent
{\bf Note}. To motivate Rule \ref{rule-binom} start with the identity
\begin{equation}
\frac{1}{A^{\alpha}} = \frac{1}{\Gamma(\alpha)} \ift x^{\alpha-1}e^{-Ax} \, dx,
\end{equation}
\noindent
and apply it to $A = a_{1} + \cdots + a_{r}$ to produce 
\begin{eqnarray}
(a_{1} + \cdots + a_{r})^{\alpha} & = & \frac{1}{\Gamma(-\alpha)} 
\ift x^{-\alpha-1} \text{exp}\left[-(a_{1}+ \cdots + a_{r})x \right] \, dx 
\nonumber \\
& = & \frac{1}{\Gamma(-\alpha)} 
\ift x^{-\alpha-1} e^{-a_{1}x} \cdots e^{-a_{r}x} \, dx. \nonumber
\end{eqnarray}
\noindent
Expanding the exponentials we obtain
\begin{equation}
(a_{1} + \cdots + a_{r})^{\alpha} \eqf 
\frac{1}{\Gamma(-\alpha)} \sum_{m_{1}} \cdots \sum_{m_{r}} 
\phi_{1,\cdots,r} a_{1}^{m_{1}} \cdots a_{r}^{m_{r}} 
\ift x^{- \alpha + m_{1} + \cdots + m_{r}-1} \, dx \nonumber 
\end{equation}
\noindent
and thus
\begin{equation}
(a_{1} + \cdots + a_{r})^{\alpha} \eqf 
\sum_{m_{1}} \cdots \sum_{m_{r}} 
\phi_{1,\cdots,r} a_{1}^{m_{1}} \cdots a_{r}^{m_{r}} 
\frac{\langle{-\alpha + m_{1} + \cdots + m_{r} \rangle} }{\Gamma(-\alpha)}.
\end{equation}
\noindent
This is Rule \ref{rule-binom}. \\

\section{Wallis' formula}
\label{sec-wallis}
\setcounter{equation}{0}

The evaluation 
\begin{equation}
J_{2,m} := \ift \frac{dx}{(1+x^{2})^{m+1}} = \frac{\pi}{2^{2m+1}} \binom{2m}{m}
\label{wallis-1}
\end{equation}
\noindent 
is historically one of the earliest closed-form expressions for a definite 
integral. The change of variables $x = \tan \theta$ converts it into its
trigonometric form
\begin{equation}
J_{2,m} := \int_{0}^{\pi/2} \cos^{2m} \theta \, d \theta = 
\frac{\pi}{2^{2m+1}} \binom{2m}{m}.
\label{wallis-2}
\end{equation}
\noindent
An elementary argument shows that $J_{2,m}$ satisfies the recurrence
\begin{equation}
J_{2,m} = \frac{2m-1}{2m}J_{2,m-1}
\end{equation}
\noindent
and then one simply checks that the right hand side of (\ref{wallis-2}) 
satisfies the same recurrence with matching initial conditions. 
A second elementary proof of (\ref{wallis-1}) is presented in \cite{sarah1}: 
using $\cos^{2} \theta = \tfrac{1}{2}(1 + \cos 2 \theta)$ one 
obtains the recurrence
\begin{equation}
J_{2,m} = 2^{-m} \sum_{i=0}^{\lfloor{ m/2 \rfloor}} \binom{m}{2i}J_{2,i},
\end{equation}
\noindent
and the inductive proof follows from the identity
\begin{equation}
\sum_{i=0}^{\lfloor{ m/2 \rfloor}} 2^{-2i} \binom{m}{2i} \binom{2i}{i} 
= 2^{-m} \binom{2m}{m}.
\end{equation}
\noindent
This can be established using automatic methods developed by H. Wilf and 
D. Zeilberger in \cite{aequalsb}.  \\

The proof of Wallis' formula by the method of brackets starts with the 
expansion of the integrand as
\begin{equation}
(1+x^{2})^{-m-1} \eqf \sum_{n_{1}} \sum_{n_{2}} \phi_{1,2} 
\frac{ \langle{m+1+n_{1}+n_{2} \rangle} }{\Gamma(m+1)}x^{2n_{2}}. 
\end{equation}
\noindent
The corresponding integral $J_{2,m}$ is assigned the bracket series 
\begin{equation}
J_{2,m} \eqf \sum_{n_{1}} \sum_{n_{2}} \phi_{1,2} 
\frac{1}{\Gamma(m+1)} 
\langle{m+1+n_{1}+n_{2} \rangle} 
\langle{2n_{2}+1 \rangle}. 
\end{equation}
\noindent
Rule \ref{rule-ass1} then shows that 
\begin{equation}
J_{2,m} =  \frac{1}{2} \frac{\Gamma(-n_{1}^{*}) \Gamma(-n_{2}^{*})}
{\Gamma(m+1)},
\end{equation}
\noindent
where $(n_{1}^{*},n_{2}^{*})$ is the solution to the linear 
system of equations
\begin{eqnarray}
m+1+n_{1}+n_{2} & = & 0, \label{system-1} \\
2n_{2}+1 & = & 0. \nonumber 
\end{eqnarray}
\noindent
Therefore $n_{1}^{*} = -(m + \tfrac{1}{2} )$ and $n_{2}^{*} = -\tfrac{1}{2}$. 
We conclude that
\begin{equation}
J_{2,m}  =  \frac{\Gamma(m + \tfrac{1}{2}) \Gamma(\tfrac{1}{2})}{2 \Gamma(m)}.
\label{wallis-3}
\end{equation}

This is exactly the right-hand side of (\ref{wallis-1}). 

\section{The integral representation of the gamma function}
\label{sec-gamma}
\setcounter{equation}{0}

The exponential in the integral 
\begin{equation}
I = \ift x^{a -1} e^{-x} \, dx 
\label{int-exp}
\end{equation}
\noindent
is expanded in power series to obtain
\begin{equation}
x^{a-1} e^{-x} = \sum_{n=0}^{\infty} \frac{(-1)^{n} x^{n+a-1}}{n!} = 
\sum_{n=0}^{\infty} \phi_{n}x^{n+a-1}.
\end{equation}
\noindent
Therefore, the integral (\ref{int-exp}) gets assigned the bracket series
\begin{equation}
I  \eqf \sum_{n} \phi_{n} \langle{a+n \rangle}.
\label{h-gamma}
\end{equation}

Rule \ref{rule-ass1}
assigns the value $\Gamma(a)$ to (\ref{h-gamma}). This is 
precisely the value of the integral:
\begin{equation}
\ift x^{a-1} e^{-x} \, dx = \Gamma(a).
\end{equation}
\noindent
Rule \ref{rule-ass1} 
was developed from this example.

\section{A Fresnel integral} \label{sec-fresnel}
\setcounter{equation}{0}

In this section we verify the evaluation of Fresnel integral
\begin{equation}
\ift \sin ( a x^{2} ) \, dx = \frac{\pi}{2 \sqrt{2a}}.
\label{fresnel-1}
\end{equation}
\noindent
The reader will find in \cite{antimirov1}  the 
standard evaluation using contour integrals and other elementary proofs in 
\cite{flanders1} and \cite{leonard1}. 

In order to apply the method of brackets, use the
hypergeometric representation 
\begin{eqnarray}
\frac{\sin z}{z} & = & 
{_{0}F_{1}} \left[ -; \tfrac{3}{2}; \, - \tfrac{z^{2}}{4} 
\right], \nonumber 
\end{eqnarray}
\noindent
that can be  written as
\begin{equation}
\sin z = \sum_{n=0}^{\infty} \phi_{n} \frac{z^{2n+1}}{\left( \tfrac{3}{2} 
\right)_{n} 4^{n} }. 
\end{equation}
\noindent
Therefore
\begin{equation}
\ift \sin ( a x^{2} ) \, dx \eqf 
\sum_{n} \phi_{n} \frac{a^{2n+1}}{\left( \tfrac{3}{2} 
\right)_{n} 4^{n} }  \langle{ 4n+3 \rangle}. 
\end{equation}
According to Rule \ref{rule-ass1}, the assignment of the
 right-hand side is obtained by
evaluating the function 
\begin{equation}
g(n) := \frac{a^{2n+1}}{\left( \tfrac{3}{2} \right)_{n} 4^{n}}
\end{equation}
at the solution of $4n^{*}+3 = 0$. Therefore the integral (\ref{fresnel-1}) 
has the value 
\begin{equation}
\tfrac{1}{4}g(-\tfrac{3}{4})
 =  \frac{a^{-1/2} \Gamma(\tfrac{3}{4}) }{\left( \tfrac{3}{2} \right)_{-3/4} 
4^{1/4} 
}, \nonumber 
\end{equation}
\noindent
where the factor $\tfrac{1}{4}$ comes from the term $4n+3$ in the bracket. 
Using $(a)_{m} = \Gamma(a+m)/\Gamma(a)$, we obtain
\begin{equation}
\left( \frac{3}{2} \right)_{-3/4} =  
\frac{2 \Gamma(\tfrac{3}{4}) }{\sqrt{\pi}}. 
\end{equation}
\noindent
We conclude that the assigned value is 
$\pi/2 \sqrt{2a}$. As expected, this is consistent with 
(\ref{fresnel-1}). \\

The method also give the evaluation of 
\begin{equation}
I = \ift x^{b-1} \sin ( ax^{c} ) \, dx. 
\label{fresnel-gen}
\end{equation}
\noindent
The change of variables $t = x^{c}$ transforms (\ref{fresnel-gen}) into 
\begin{equation}
I = \frac{1}{c} \ift t^{b/c -1} \sin(at) \, dt,
\end{equation}
\noindent
and this is formula $3.761.4$ in \cite{gr} with value
\begin{equation}
\ift x^{b-1} \sin(ax^{c}) \, dx  
= \frac{\Gamma(b/c)}{ca^{b/c}} \sin \left( \frac{ \pi b }{2c} \right).
\label{modi-2}
\end{equation}

To verify this result by the method of brackets, start with the 
expansion
\begin{equation}
x^{b-1} \sin( ax^{c} ) =  \sum_{n =0}^{\infty} 
\phi_{n} \frac{a^{2n+1}}{\left( \tfrac{3}{2} \right)_{n} 2^{2n}} 
x^{2nc + c + b -1 }
\end{equation}
\noindent
and associate to it the bracket series
\begin{equation}
\ift x^{b-1} \sin ( ax^{c} ) \, dx
 \eqf \sum_{n} 
\phi_{n} \frac{a^{2n+1}}{\left( \tfrac{3}{2} \right)_{n} 2^{2n}} 
\langle{ 2nc + c + b \rangle}.
\end{equation}
\noindent 
Apply Rule \ref{rule-ass1} to obtain
\begin{equation}
I   =  \frac{1}{2c} \frac{a^{2n_{*}+1}}{\left( \tfrac{3}{2} \right)_{n^{*}} 
2^{2n^{*}} } \Gamma(-n^{*}),
\label{value-11}
\end{equation}
\noindent
where $n^{*}$ solve $2nc + b + c = 0$; that is, $n^{*} = -1/2 - b/2c$. 
Then (\ref{value-11}) yields
\begin{equation}
I  =  \frac{\Gamma(\tfrac{3}{2}) 
2^{b/c}}{c a^{b/c} \Gamma(1 - \tfrac{b}{2c}) } 
\Gamma \left( \tfrac{1}{2} + \tfrac{b}{2c} \right). 
\label{modi-1}
\end{equation}
\noindent
with $x = b/2c$ to transform (\ref{modi-1}) into (\ref{modi-2}). 
To transform (\ref{modi-1}) into (\ref{modi-2}), simplify 
(\ref{modi-1}) using the reflection formula 
\begin{equation}
\Gamma(x) \Gamma(1-x) = \frac{\pi}{\sin \pi x},
\end{equation}
\noindent
and the duplication formula
\begin{equation}
\Gamma( x + \tfrac{1}{2} ) =  \frac{\Gamma(2x) \sqrt{\pi}}{\Gamma(x) 
2^{2x-1}},
\end{equation}
\noindent
with $x = b/2c$. 

\medskip

\noindent
{\bf Note}. The method developed by Flanders \cite{flanders1} is based on 
showing that
\begin{equation}
F(t) := \ift e^{-tx^{2}} \cos x^{2} \, dx \, \, \text{ and } \, \, 
G(t) := \ift e^{-tx^{2}} \sin x^{2} \, dx 
\label{fresnel-00}
\end{equation}
\noindent
satisfy the functional equation 
\begin{equation}
F^{2}(t) - G^{2}(t) = 2F(t)G(t) = \frac{\pi}{4(1+t^{2})}.
\end{equation}
\noindent
The latter  can be solved to obtain the values
\begin{equation}
F(t) = \sqrt{\frac{\pi}{8}} \sqrt{ \frac{\sqrt{1+t^{2}}+t}{1+t^{2}}} 
\text{ and }
G(t) = \sqrt{\frac{\pi}{8}} \sqrt{ \frac{\sqrt{1+t^{2}}-t}{1+t^{2}}}.
\end{equation}
\noindent
A second elementary proof was obtained by Leonard \cite{leonard1}. 
Converting (\ref{fresnel-00}) into the Laplace 
transform of $\cos x/2 \sqrt{x}$ and $\sin x/2 \sqrt{x}$
respectively, he shows that 
\begin{equation}
F(t) = \frac{1}{\sqrt{\pi}} \ift \frac{u^{2}+t}{1+(u^{2}+t)^{2}} \, du 
\,\, \text{ and } \, \, 
G(t) = \frac{1}{\sqrt{\pi}} \ift \frac{du}{1+(u^{2}+t)^{2}}. 
\end{equation}
\noindent
The  evaluation of these integrals described in \cite{leonard1}, is 
elementary but long. A shorter argument follows from the formula
\begin{equation}
f(a) := \ift \frac{dx}{x^{4} + 2ax^{2}+1} = \frac{\pi}{2 \sqrt{2(1+a)}}. 
\label{quartic-00}
\end{equation}
\noindent
Indeed, the values
\begin{equation}
G(t) = (1+t^{2})^{-3/4} f \left( \frac{t}{\sqrt{t^{2}+1}} \right) 
\text{ and } 
F(t) = (1+t^{2})^{-1/4} G(t/\sqrt{1+t^{2}}) + tG(t),
\nonumber
\end{equation}
\noindent
follow from (\ref{quartic-00}) by a change of 
variable $v = (1+t^{2})^{1/4}u$.  The evaluation of the 
quartic integral (\ref{quartic-00})
by the method of brackets 
is discussed in detail in Section \ref{sec-quartic}.

\section{An integral of beta type} \label{sec-beta}
\setcounter{equation}{0}

In this section we present the evaluation of 
\begin{equation}
I = \ift \frac{x^{a} \, dx}{(E + Fx^{b})^{c}}.
\label{beta-1}
\end{equation}
\noindent
The change of variables $x = C^{1/b}t^{1/b}$, with $C = E/F$,  yields
\begin{equation}
I = \frac{C^{u}}{bE^{c}} \ift \frac{t^{u-1} \, dt}{(1+t)^{c}}
\end{equation}
\noindent
where  $u = (a+1)/b$. The new 
integral evaluates as $B( c - u, u )$ where $B(x,y)$ is the classical beta 
function; see \cite{gr}, formula $8.380.3$. We conclude that 
\begin{equation}
I = \frac{C^{u}}{bE^{c}} B ( c -u , u).
\label{value-g}
\end{equation}

To evaluate this integral by the method of brackets, the integrand 
$(E + Fx^{b})^{-c}$ is expanded as
\begin{equation}
\sum_{n_{1}} \sum_{n_{2}} \phi_{1,2} E^{n_{1}} F^{n_{2}} x^{bn_{2}} 
\frac{\langle{ c + n_{1}+n_{2} \rangle}}{\Gamma(c)}. 
\end{equation}
\noindent
Replacing in (\ref{beta-1}) we obtain
\begin{eqnarray}
I & \eqf & \sum_{n_{1}} \sum_{n_{2}} \phi_{1,2} E^{n_{1}}F^{n_{2}} 
\frac{\langle{ c + n_{1} + n_{2} \rangle} }{\Gamma(c)} 
\ift x^{a + bn_{2}+1-1} \, dx \nonumber \\
 & \eqf & \sum_{n_{1}} \sum_{n_{2}} \phi_{1,2} E^{n_{1}}F^{n_{2}} 
\frac{1}{\Gamma(c)} \langle{ c + n_{1} + n_{2} \rangle} 
\langle{a+ bn_{2} + 1 \rangle}. \nonumber 
\end{eqnarray}

To obtain the value assigned to the two dimensional sum, solve
\begin{eqnarray}
c + n_{1} + n_{2} & = & 0 \nonumber \\
a + bn_{2} + 1 & = & 0, \nonumber 
\end{eqnarray}
\noindent
to produce the solution $n_{1}^{*} = \tfrac{a+1}{b} - c$ and $n_{2}^{*} = 
- \frac{a+1}{b}$. Therefore
\begin{equation}
I  =  \frac{1}{b \Gamma(c)} 
E^{n_{1}^{*}} F^{n_{2}^{*}} \Gamma( - n_{1}^{*}) \Gamma(-n_{2}^{*}),
\end{equation}
\noindent
and this reduces to the value in (\ref{value-g}).

\section{A combination of powers and exponentials } \label{sec-comb1}
\setcounter{equation}{0}

In this section we employ the method of brackets and evaluate the integral
\begin{equation}
I := \ift \frac{x^{\alpha -1 } \, dx }{\left( A + B \text{exp}(C x^{\beta}) 
\right)^{\gamma}}, 
\label{comb-exp1}
\end{equation}
\noindent
with $\alpha, \, \beta, \, \gamma, \, A, \, B, \, C \in \mathbb{R}$. 
To evaluate this integral we consider  the bracket series
\begin{equation}
\left( A + B \text{exp}(C x^{\beta}) \right)^{-\gamma} \eqf 
\sum_{n_{1},n_{2}} A^{n_{1}} B^{n_{2}} \text{exp}(C n_{2}x^{\beta}) 
\frac{\langle{\gamma + n_{1} + n_{2} \rangle}}{\Gamma(\gamma)}. 
\end{equation}
\noindent
The exponential function is expanded as
\begin{eqnarray}
\text{exp}(C n_{2} x^{\beta}) & = & 
\sum_{n_{3}=0}^{\infty} \frac{C^{n_{3}} n_{2}^{n_{3}}}{\Gamma(n_{3}+1)} 
x^{\beta n_{3}} \nonumber \\
& =  &  \sum_{n_{3}=0}^{\infty} C^{n_{3}} (-n_{2})^{n_{3}} \phi_{n_{3}} 
x^{n_{3}}. \nonumber
\end{eqnarray}
\noindent
Therefore, the integral  (\ref{comb-exp1}) is assigned the bracket series

\begin{equation}
I \eqf \sum_{n_{1},n_{2},n_{3}} \phi_{1,2,3} 
\frac{A^{n_{1}} B^{n_{2}} C^{n_{3}} (-n_{2})^{n_{3}} 
\langle{ \alpha + \beta n_{3} \rangle} 
\langle{ \gamma + n_{1} + n_{2} \rangle}}{\Gamma(\gamma)}.
\nonumber
\end{equation}

\noindent
The vanishing of the two brackets leads to the system
\begin{eqnarray}
\alpha + \beta n_{3} & = & 0 \nonumber \\
\gamma + n_{1} + n_{2} & = & 0, \nonumber 
\end{eqnarray}
\noindent
and we have to choose a free parameter 
between $n_{1}$ and $n_{2}$. Observe that
$n_{3} = - \alpha/ \beta$ is determined by the method.  \\

\noindent
{\bf Choice 1}: take $n_{2}$ to be free. Then $n_{1}^{*} = -\gamma - n_{2}$ and 
$n_{3}^{*} = - \alpha/\beta$. This leads to
\begin{equation}
I  =  \sum_{n_{2}=0}^{\infty} \frac{B^{n_{2}}  \Gamma(\alpha/\beta) 
\Gamma(\gamma + n_{2})}{A^{\gamma + n_{2}} C^{\alpha/\beta} 
\beta \Gamma(\gamma) (-n_{2})^{\alpha/\beta} }. \noindent
\end{equation}
\noindent
This is impossible due to the presence of the term $n_{2}^{\alpha/\beta}$ 
leading to a divergent series. These divergent series are discarded. \\

\noindent
{\bf Choice 2}: take $n_{1}$ as the free variable. Then 
$n_{3}^{*} = -\alpha/\beta$ and $n_{2}^{*} = -\gamma - n_{1}$. This time we 
obtain
\begin{equation}
I  = \frac{\Gamma(\alpha/\beta)}{\Gamma(\gamma)} 
\frac{1}{B^{\gamma} C^{\alpha/\beta} \, \beta} 
\sum_{n_{1}=0}^{\infty}  (-1)^{n_{1}}
\frac{\Gamma(\gamma+n_{1})}{\Gamma(1+n_{1})}
\frac{(A/B)^{n_{1}}}{(\gamma + n_{1})^{\alpha/\beta}}. 
\end{equation}
\noindent
This formula cannot be expressed in term of more elementary special 
functions. 

In the special case $\gamma =1$ we obtain 
\begin{equation}
I  = - \frac{\Gamma(\nu)}{a \beta c^{\nu}} \text{PolyLog}(\nu,-a/b). 
\end{equation}
\noindent
with  $\nu = \alpha/\beta$. The
polylogarithm function appearing here is defined by 
\begin{equation}
\text{PolyLog}(z,k) := \sum_{n=1}^{\infty} \frac{z^{n}}{n^{k}}.
\end{equation}
\noindent
Specializing to $A=B=C=\alpha=\gamma =1$ and $\beta = 2$ we obtain
\begin{equation}
\ift \frac{dx}{1+e^{x^{2}}}  = - \frac{\sqrt{\pi}}{2} (\sqrt{2}-1) 
\zeta \left( \tfrac{1}{2} \right). 
\end{equation}
\noindent
Of course, this 
integral can be evaluated by simply expanding the integrand as a
geometric series.

\section{The Mellin transform of a quadratic exponential}
 \label{sec-mellin}
\setcounter{equation}{0}

The Mellin transform of a function $f(x)$ is defined by 
\begin{equation}
\mathcal{M}(f)(s) := \ift x^{s-1} f(x) \, dx.
\end{equation}
\noindent
Many of the integrals appearing in \cite{gr} are of this type. For example, 
$3.462.1$ states that
\begin{equation}
\mathcal{M}\left( e^{-\beta x^{2} - \gamma x} \right)(s)  = 
\ift x^{s-1} e^{- \beta x^{2} - \gamma x} \, dx = 
(2 \beta)^{-s/2} \Gamma(s) e^{\gamma^{2}/(8 \beta)} D_{-s} 
\left( \frac{\gamma}{\sqrt{2 \beta}} \right). 
\label{form-hyper1}
\end{equation}
\noindent
Here $D_{p}(z)$ is the parabolic cylinder function defined 
by (formula $9.240$ in 
\cite{gr})
\begin{equation}
D_{p}(z) = 2^{p/2} e^{-z^{2}/4} 
\left( \frac{\sqrt{\pi}}{\Gamma((1-p)/2)} {_{1}F_{1}}  \left( - \frac{p}{2}, 
\frac{1}{2}; \frac{z^{2}}{2} \right) - 
\frac{\sqrt{2 \pi}z}{\Gamma(-p/2)} {_{1}F_{1}} \left( \frac{1-p}{2}, 
\frac{3}{2}; \frac{z^{2}}{2} \right) 
\right).
\nonumber
\end{equation}

A direct application of the method of brackets gives
\begin{equation}
\ift x^{s-1} e^{- \beta x^{2} - \gamma x} \, dx  \eqf 
\sum_{n_{1}} \sum_{n_{2}} \phi_{1,2} \beta^{n_{1}} \gamma^{n_{2}} 
\langle{ s + 2n_{1} + n_{2} \rangle}.
\end{equation}

The equation $s+2n_{1} + n_{2}=0$ gives two choices for a 
free index. Taking $n_{2}^{*} = -2n_{1}-s$ leads 
to the series 
\begin{eqnarray}
\sum_{n_{1}=0}^{\infty} \frac{1}{\Gamma(n_{1}+1)}  
\left( - \frac{\beta}{\gamma^{2}} \right)^{n_{1}} (s)_{2n_{1}} & = & 
\sum_{n_{1}=0}^{\infty} \frac{1}{\Gamma(n_{1}+1)}  
\left( - \frac{4 \beta}{\gamma^{2}} \right)^{n_{1}}  
\left( \frac{s}{2} \right)_{n_{1}} 
\left( \frac{s+1}{2} \right)_{n_{1}}  \nonumber \\
 & = &  
{_{2}F_{0}} \left( \frac{s}{2}, \frac{s+1}{2} \Big{|} - 
\frac{4 \beta}{\gamma^{2}} \right). 
\nonumber
\end{eqnarray}
\noindent
This choice of a free 
index is excluded because the resulting series diverges.  The second choice 
is $n_{1}^{*} = -n_{2}/2-s/2$
and this yields the series
\begin{equation}
\frac{1}{2 \beta^{s/2}} \sum_{n_{2}=0}^{\infty} 
\frac{\rho^{n_{2}}}{\Gamma(n_{2}+1)}
\Gamma \left( \frac{n_{2}}{2} + \frac{s}{2} \right),
\label{mess-11}
\end{equation}
\noindent
where $\rho = - \gamma/\sqrt{\beta}$. 
To write (\ref{mess-11}) in 
hypergeometric form we separate it into two sums according to
the parity of $n_{2}$ and obtain
\begin{equation}
\frac{1}{2 \beta^{s/2}} 
\left( 
\Gamma \left( \frac{s}{2} \right) 
\sum_{n=0}^{\infty} \frac{\rho^{2n}}{(1)_{2n}} \left( \frac{s}{2} \right)_{n}+
\Gamma \left( \frac{1+s}{2} \right) 
\sum_{n=0}^{\infty} \frac{\rho^{2n+1}}{(2)_{2n}} 
\left( \frac{1+s}{2} \right)_{n} 
\right). 
\nonumber
\end{equation}
\noindent
The identity 
\begin{equation}
(a)_{2n} = 4^{n} \left( \frac{a}{2} \right)_{n} 
\left( \frac{a+1}{2} \right)_{n}
\end{equation}
\noindent
gives the final representation of the sum as 
\begin{equation}
\frac{1}{2 \beta^{s/2}} 
\left[
\Gamma \left( \frac{s}{2} \right) 
{_{1}F_{1}} \left( \frac{s}{2}, \frac{1}{2}; \frac{1}{2}\rho^{2} \right) +
\rho \Gamma \left( \frac{1+s}{2} \right) 
{_{1}F_{1}} \left( \frac{1+s}{2}, \frac{3}{2}; \frac{1}{2}\rho^{2} \right)
\right].
\end{equation}
\noindent
This is (\ref{form-hyper1}).  \\

The special case $s = 1$ gives
\begin{equation}
\ift e^{-\beta x^{2} - \gamma x} \, dx  =  
\tfrac{1}{2\sqrt{\beta}} \left[ \Gamma(\tfrac{1}{2})  \, \, 
{_{1}F_{1}}\left( \tfrac{1}{2}; \tfrac{1}{2}; \tfrac{1}{2} \rho^{2} \right) 
+ \rho \, 
{_{1}F_{1}}\left( 1; \tfrac{3}{2}; \tfrac{1}{2} \rho^{2} \right) \right].
\end{equation}
The first hypergeometric sum evaluates to $e^{\gamma^{2}/4 \beta}$ and 
using the
representation of the {\em error function}
\begin{equation}
\text{erf}(x) = \frac{2}{\sqrt{\pi}} \int_{0}^{x} e^{-t^{2}} \, dt
\end{equation}
\noindent
as 
\begin{equation}
\text{erf}(x) = \frac{2x}{\sqrt{\pi}} e^{-x^{2}} 
{_{1}F_{1}}\left( 1; \tfrac{3}{2}; x^{2} \right),
\end{equation}
(given as $8.253.1$ in \cite{gr}) we find the value of the second 
hypergeometric sum. The conclusion is that 
\begin{equation}
\ift e^{-\beta x^{2} - \gamma x} \, dx  =   \frac{\sqrt{\pi}}{2 \beta} 
\text{exp} \left( \frac{\gamma^{2}}{4 \beta} \right) 
\left( 1 - \text{erf} \left( \frac{\gamma}{2 \sqrt{\beta}} \right) \right). 
\end{equation}
\noindent
This can be checked directly by completing the square in the integrand.

\section{A multidimensional integral from Gradshteyn and Ryzhik} 
\label{sec-multi}
\setcounter{equation}{0}

The method of brackets can also be used to evaluate some multidimensional 
integrals. Consider the following integral
\begin{equation}
I_{n} := \ift \ift \cdots \ift 
\frac{x_{1}^{p_{1}-1} x_{2}^{p_{2}-1} \cdots x_{n}^{p_{n}-1} 
\, \, dx_{1}dx_{2} \cdots dx_{n}}
{\left( 1 + (r_{1}x_{1})^{q_{1}} + \cdots + (r_{n}x_{n})^{q_{n}} \right)^{s}},
\end{equation}
\noindent
which appears as $4.638.3$ in \cite{gr} with an incorrect 
evaluation.  \\

The first step in the evaluation of $I_{n}$ is to expand the denominator of 
the integrand using Rule \ref{rule-binom} as
\begin{equation}
\frac{1}
{\left( 1 + (r_{1}x_{1})^{q_{1}} + \cdots + (r_{n}x_{n})^{q_{n}} \right)^{s}}
\eqf 
\sum_{k_{0}, k_{1}, \cdots, k_{n}} 
\phi_{0,\cdots,n} \prod_{j=1}^{n} (r_{j}x_{j})^{q_{j}k_{j}} 
\frac{\langle{s+ k_{0} + \cdots + k_{n} \rangle} }{\Gamma(s)}.
\nonumber
\end{equation}
\noindent
Next the integral is assigned the value
\begin{equation}
I_{n}  \eqf 
\sum_{k_{0}, k_{1}, \cdots, k_{n}} 
\phi_{0,\cdots,n} \prod_{j=1}^{n} (r_{j}x_{j})^{q_{j}k_{j}} 
\frac{\langle{s+ k_{0} + \cdots + k_{n} \rangle} }{\Gamma(s)}
\prod_{j=1}^{n} \langle{ p_{j} + q_{j}k_{j} \rangle}.
\end{equation}
The evaluation of this bracket sum involves the values 
\begin{equation}
k_{0} = -s + \sum_{j=1}^{n} \frac{p_{j}}{q_{j}} \text{ and }
k_{j} = - \frac{p_{j}}{q_{j}} \text{ for } 1 \leq j \leq n. 
\end{equation}
\noindent
We conclude that
\begin{equation}
I_{n}  =  
\frac{1}{\Gamma(s)} 
\Gamma \left( s - \sum_{j=1}^{n} \frac{p_{j}}{q_{j}} \right)
\prod_{j=1}^{n}  \frac{\Gamma \left( \tfrac{p_{j}}{q_{j}} \right)}
{q_{j}r_{j}^{p_{j}} }.
\end{equation}
The table \cite{gr} has the exponents of $r_{j}$ written as $p_{j}q_{j}$ 
instead of $p_{j}$. This has now been corrected.

\section{An example involving Bessel functions} \label{sec-bessel}
\setcounter{equation}{0}

The Bessel function $J_{\nu}(x)$ is defined by the series
\begin{equation}
J_{\nu}(x) = \frac{1}{2^{\nu}} \sum_{k=0}^{\infty} (-1)^{k} 
\frac{z^{2k+\nu}}{2^{2k} k! \Gamma(\nu + k + 1)},
\end{equation}
\noindent
and it admits the hypergeometric representation
\begin{equation}
J_{\nu}(x) = \frac{x^{\nu}}{2^{\nu} \, \Gamma(1+ \nu)} \, 
{_{0}F_{1}} 
\left(
\begin{matrix}
 -  \\
1 + \nu  
\end{matrix}
\Big{|} \frac{-x^{2}}{4} 
\right).
\end{equation}
\noindent
The method of brackets will now be employed to evaluate the integral
\begin{equation}
I := \ift x^{-\lambda} J_{\nu}( \alpha x ) J_{\mu}( \beta x) \, dx.
\label{bessel-1}
\end{equation}
\noindent
Three integrals of this type form Section $6.574$ of \cite{gr}.  \\

Replacing the hypergeometric form  in the integral, we have
\begin{eqnarray}
I & \eqf & \frac{ \left( \tfrac{\alpha}{2} \right)^{\nu} 
              \left( \tfrac{\beta}{2} \right)^{\mu} }
{\Gamma(\nu+1) \Gamma(\mu+1)}  \nonumber \\
& \times & \ift  \sum_{n_{1}, n_{2}} 
\phi_{1,2} \frac{\alpha^{2n_{1}} \, \beta^{2n_{2}} }
{4^{n_{1}+n_{2}} \, (\nu+1)_{n_{1}} \, (\mu+1)_{n_{2}}} 
x^{2n_{1}+2n_{2}-\lambda +\nu + \mu} \, dx.\nonumber
\end{eqnarray}
\noindent
Therefore, the bracket series associated to the integral (\ref{bessel-1}) 
becomes 
\begin{eqnarray}
I & \eqf & \frac{ 2^{-\nu-\mu} \alpha^{\nu} 
              \beta^{\mu}}
{\Gamma(\nu+1) \Gamma(\mu+1)}  \nonumber \\
& \times & \sum_{n_{1}} \sum_{n_{2}} 
\frac{\phi_{1,2}}{ 4^{n_{1}+ n_{2}}} \frac{\alpha^{2n_{1}} \, \beta^{2n_{2}} }
{(\nu+1)_{n_{1}} \, (\mu+1)_{n_{2}}} 
\langle{ 2n_{1}+2n_{2}-\lambda +\nu + \mu + 1 \rangle}.\nonumber
\end{eqnarray}

The vanishing of the brackets yields the value 
$n_{1}^{*} = \frac{1}{2}(\lambda- \nu - \mu -1)
-n_{2}$ and it follows that
\begin{equation}
I  =  \frac{2^{-\nu -\mu}}{\Gamma(\nu+1) \Gamma(\mu+1)} 
\sum_{n_{2}=0}^{\infty} \frac{\phi_{2}}{ 4^{n_{1}^{*} +n_{2}}} 
\frac{\alpha^{2 n_{1}^{*}} \beta^{2n_{2}} }{(\nu+1)_{n_{1}^{*}} 
(\mu+1)_{n_{2}}}
\frac{\Gamma(-n_{1}^{*} )}{2}. \nonumber
\end{equation}

Writing the Pochhammer symbol $(\nu+1)_{n_{1}^{*}}$ in terms of the 
gamma function we obtain 
\begin{eqnarray}
I &  = & \frac{\beta^{\mu} \alpha^{\lambda - \mu -1} }
{2^{\lambda} \Gamma(\mu+ +1)} \nonumber \\
& \times & \sum_{n_{2}=0}^{\infty} 
\frac{(-1)^{n_{2}} }{\Gamma(n_{2}+1)} 
\frac{( \beta^{2}/\alpha^{2})^{n_{2}} }
{\Gamma( \nu + 1 + \tfrac{1}{2}(\lambda - \nu -\mu-1) - n_{2} ) } 
\frac{\Gamma( \tfrac{1}{2}(\nu+\mu-\lambda+1) + n_{2})}{(\mu+1)_{n_{2}}}. 
\nonumber 
\end{eqnarray}
\noindent
In order to write this in hypergeometric terms, we start with 
\begin{eqnarray}
I &  = & \frac{\beta^{\mu} \alpha^{\lambda - \mu -1} }
{2^{\lambda} \Gamma(\mu+ +1)} \nonumber \\
 & \times & 
\sum_{n_{2}=0}^{\infty} 
(-1)^{n_{2}} 
\frac{ ( \tfrac{1}{2} ( \nu + \mu - \lambda +1) )_{n_{2}}  
(\beta^{2}/\alpha_{2})^{n_{2}} }
{( \tfrac{1}{2}(\lambda + \nu - \mu +1))_{-n_{2}} \, (\mu+1)_{n_{2}} 
\Gamma(n_{2}+1)},
\nonumber
\end{eqnarray}
\noindent
and use the identity 
\begin{equation}
(c)_{-n} = \frac{(-1)^{n}}{(1-c)_{n}}, 
\end{equation}
\noindent
to  obtain
\begin{eqnarray}
I &  =  & \frac{\beta^{\mu} \alpha^{\lambda-\mu-1}}{2^{\lambda}} 
\frac{\Gamma( \tfrac{1}{2}(\nu+\mu-\lambda+1)}
{\Gamma(\mu+1) \Gamma( \tfrac{1}{2} (\lambda + \nu - \mu+1) )} \nonumber \\
& \times & 
\sum_{n_{2}=0}^{\infty} 
( \tfrac{1}{2} (1 - \lambda-\nu+\mu) )_{n_{2}} 
( \tfrac{1}{2} (\nu+\mu - \lambda + 1) )_{n_{2}} 
\frac{1}{(\mu+1)_{n_{2}} \, \Gamma(n_{2}+1)} 
\left( \frac{\beta^{2}}{\alpha^{2}} \right)^{n_{2}}, 
\nonumber 
\end{eqnarray}
\noindent
that can be written as 
\begin{eqnarray}
I &  =  & \frac{\beta^{\mu} \alpha^{\lambda-\mu-1}}{2^{\lambda}} 
\frac{\Gamma( \tfrac{1}{2}(\nu+\mu-\lambda+1)) }
{\Gamma(\mu+1) \Gamma(\tfrac{1}{2}(\lambda + \nu-\mu+1)) } \nonumber \\
& \times & {_{2}F_{1}} 
\left(
\begin{matrix}
\tfrac{1}{2}(1 - \lambda -\nu + \mu) &  & \tfrac{1}{2}(\nu+\mu-\lambda+1)  \\
 & \mu+1 & 
\end{matrix}
\Big{|} \frac{\beta^{2}}{\alpha^{2}} 
\right). 
\nonumber
\end{eqnarray}

This solution is valid for $|\beta^{2}/\alpha^{2} | < 1$ and it corresponds to 
formula $6.574.3$ in \cite{gr}. The table contains an error in this formula,
the power of $\beta$ is written as $\nu$ instead of $\mu$. To obtain a 
formula valid for $|\beta^{2}/\alpha^{2}| > 1$ we could proceed as before 
and obtain $6.574.1$ in \cite{gr}. Alternatively exchange 
$(\nu, \alpha)$ by $(\mu, \beta)$ and use the formula developed above. 

\section{A new evaluation of a quartic integral} \label{sec-quartic}
\setcounter{equation}{0}

The integral 
\begin{equation}
N_{0,4}(a;m) := \ift \frac{dx}{(x^{4} + 2ax^{2} + 1)^{m+1}}
\end{equation}
\noindent 
is given by 
\begin{equation}
N_{0,4}(a,m) = \frac{\pi}{2} \frac{P_{m}(a) }{[2(a+1)]^{m+\tfrac{1}{2}}},
\label{value-quartic}
\end{equation}
\noindent
where $P_{m}$ is the polynomial
\begin{equation}
P_{m}(a)  = \sum_{l=0}^{m} d_{l,m}a^{l},
\end{equation}
\noindent
with coefficients
\begin{equation}
d_{l,m} = 2^{-2m} \sum_{k=l}^{m} 2^{k} \binom{2m-2k}{m-k} \binom{m+k}{m}
\binom{k}{l}.
\end{equation}
The sequence 
$\{ d_{l,m}: \, 0 \leq l \leq m \}$ have remarkable arithmetical and 
combinatorial properties
\cite{manna-moll-survey}.  \\

The reader will find in \cite{amram} a survey of the many different proofs
of (\ref{value-quartic}) available in the literature. One of these proofs
follows from  the hypergeometric representation 
\begin{equation}
N_{0,4}(a,m) = 2^{m - \tfrac{1}{2}} (a+1)^{-m - \tfrac{1}{2} } 
B \left( 2m + \tfrac{3}{2} \right) 
{_{2}F_{1}} 
\left(
\begin{matrix}
-m &  & m+1  \\
 & m + \tfrac{3}{2}  &   
\end{matrix}
\Big{|} \frac{1-a}{2} 
\right).
\end{equation}

New proofs of this evaluation keep on appearing. For instance, the survey 
\cite{amram} does not include the recent automatic proof by 
C. Koutschan and V. Levandovskyy \cite{koutschan1}. 
The goal of this section is to provide yet another
proof of the identity (\ref{value-quartic}) using the method of brackets.

The bracket series for $I \equiv N_{0,4}(a,m)$ is formed by the usual 
procedure. The result is 
\begin{equation}
\label{N-bracket}
I \eqf \frac{1}{\Gamma(m+1)} \sum_{n_{1},n_{2},n_{3}} 
\phi_{1,2,3} (2a)^{n_{2}} 
\langle{ 4n_{1}+2n_{2}+1 \rangle} 
\langle{ m+1 + n_{1}+n_{2}+n_{3} \rangle}.
\end{equation}
\noindent 

The expression (\ref{N-bracket}) contains 
two brackets and three indices. Therefore the final result will 
be a single series on the free index. We employ the following notation: 
$I$ is the original bracket series, the symbol  $I_{j}$ denotes the series
$I$ after eliminating the index $n_{j}$. Similarly $I_{i,j}$ denotes the 
series $I$ after first eliminating $n_{i}$ (to produce $I_{i}$) and then 
eliminating $n_{j}$.  \\

\noindent
{\bf Case 1}: $n_{3}$ is the free index.  Eliminate first $n_{1}$ from 
the bracket $\langle{ 4n_{1} + 2n_{2}+1 \rangle}$ to obtain 
$n_{1}^{*} = - \tfrac{1}{2}n_{2} - \tfrac{1}{4}$. The resulting bracket 
series is 
\begin{equation}
I_{1} \eqf  \sum_{n_{2},n_{3}} \phi_{2,3} 
\frac{(2a)^{n_{2}} \Gamma(\tfrac{1}{2}n_{2} + \tfrac{1}{4}) }{4 \Gamma(m+1) } 
\langle{ m + \tfrac{3}{4} + \tfrac{1}{2}n_{2} + n_{3} \rangle}.
\end{equation}
\noindent
The next step is to eliminate $n_{2}$ to get $n_{2}^{*} = -2m- \tfrac{3}{2} 
- 2n_{3}$ and obtain
\begin{equation}
I_{1,2}  =  \frac{1}{2 \Gamma(m+1) (2a)^{2m+3/2}} 
\sum_{n_{3}=0}^{\infty} \frac{\phi_{3}}{(2a)^{n_{3}}} 
\Gamma( -m-\tfrac{1}{2}-n_{3}) \Gamma( 2m + \tfrac{3}{2} + 2n_{3}).
\end{equation}

In order to simplify these expressions, we employ
\begin{equation}
\Gamma(x+m)  =  (x)_{m} \Gamma(x), \, 
\Gamma(x-m)  =  (-1)^{m}\Gamma(x)/(1-x)_{m}
\label{reduce-1} 
\end{equation}
\noindent
and
\begin{equation}
(x)_{2m} =  2^{2m} 
\left(\tfrac{1}{2}x \right)_{m} \left( \tfrac{1}{2}(x+1) \right)_{m},
\label{reduce-2}
\end{equation}
\noindent 
for $x \in \mathbb{R}$ and $m \in \mathbb{N}$. We obtain
\begin{equation}
\Gamma( -m - \tfrac{1}{2} - n_{3}) = 
\frac{(-1)^{n_{3}}\Gamma(-\tfrac{1}{2}-m)}{( \tfrac{3}{2} + m)_{n_{3}}}
\nonumber
\end{equation}
\noindent
and 
\begin{equation}
\Gamma(2m+ \tfrac{3}{2} + 2n_{3}) = 
\Gamma(2m+ \tfrac{3}{2}) (m + \tfrac{3}{4})_{n_{3}} 
(m + \tfrac{5}{4})_{n_{3}} 2^{2n_{3}}. 
\nonumber
\end{equation}
\noindent
These  yield
\begin{equation}
I_{1,2} =  
\frac{\Gamma(-\tfrac{1}{2} -m) \Gamma( 2m + \tfrac{3}{2})}
{2 \Gamma(m+1) (2a)^{2m+ 3/2}} 
\sum_{n_{3}=0}^{\infty} 
\frac{(m + 3/4)_{n_{3}} \, (m + 5/4)_{n_{3}} }{(m + 3/2)_{n_{3}} n_{3}!}
a^{-2n_{3}},
\end{equation}
\noindent
or
\begin{equation}
I_{1,2}  =  
\frac{\Gamma(-\tfrac{1}{2} -m) \Gamma( 2m + \tfrac{3}{2})}
{2 \Gamma(m+1) (2a)^{2m+ 3/2}} 
\, \, {_{2}F_{1}} 
\left(
\begin{matrix}
m+ \tfrac{3}{4}  &  & m + \tfrac{5}{4}   \\
 & m + \tfrac{3}{2}  &   
\end{matrix}
\, \Big{|} \frac{1}{a^{2}} 
\right).
\end{equation}

\noindent
{\bf Note}. The reader 
can check that $I_{1,2} = I_{2,1}$, so the value of the sum for 
the quartic integral does not depend on the order in which the indices 
$n_{1}$ and $n_{2}$ are eliminated. The reader can also verify that this 
occurs in the next two cases described below; that is, $I_{1,3} = I_{3,1}$ 
and $I_{2,3} = I_{3,2}$.

\medskip

\noindent
{\bf Case 2}: $n_{1}$ is the free index.  A similar argument yields
\begin{equation}
I_{2,3}  =  
\frac{\Gamma(m + \tfrac{1}{2}) \Gamma( \tfrac{1}{2})}
{2 \Gamma(m+1) (2a)^{1/2}} 
\, \, {_{2}F_{1}} 
\left(
\begin{matrix}
\tfrac{1}{4}  &  & \tfrac{3}{4}   \\
 & \tfrac{1}{2} - m  &   
\end{matrix}
\, \Big{|} \frac{1}{a^{2}} 
\right).
\end{equation}

\medskip

\noindent
{\bf Case 3}: $n_{2}$ is the free index.  Eliminate $n_{1}$ from the 
bracket series (\ref{N-bracket})  to produce
\begin{equation}
I_{1}  \eqf   \sum_{n_{2},n_{3}} \phi_{2,3} 
\frac{(2a)^{n_{2}} \Gamma(\tfrac{1}{2}n_{2} + \tfrac{1}{4}) }{4 \Gamma(m+1) } 
\langle{ m + \tfrac{3}{4} + \tfrac{1}{2}n_{2} + n_{3} \rangle}, 
\end{equation}
\noindent
and now eliminate $n_{3}$ to obtain $n_{3}^{*} = -m - \tfrac{3}{4} - 
\tfrac{1}{2}n_{2}$. This yields
\begin{equation}
I_{1,3}  =  
\frac{1}{4 \Gamma(m+1)} \sum_{n_{2}=0}^{\infty} (-1)^{n_{2}} 
\frac{(2a)^{n_{2}}}{n_{2}!} \Gamma( \tfrac{1}{2}n_{2} + \tfrac{1}{4} ) 
\Gamma( m+ \tfrac{3}{4} + \tfrac{1}{2}n_{2}). 
\end{equation}

In order to obtain a hypergeometric representations of these 
expressions, we separate the last  series 
according to the parity of $n_{2}$:

\begin{eqnarray}
I_{1,3} &  = & 
\frac{1}{4 \Gamma(m+1)} 
\sum_{n_{2}=0}^{\infty} \frac{(2a)^{2n_{2}}}{(2n_{2})!} \, 
\Gamma( n_{2} + \tfrac{1}{4}) 
\Gamma( n_{2} + m + \tfrac{3}{4})  \nonumber \\
& - & \frac{1}{4 \Gamma(m+1)} 
\sum_{n_{2}=0}^{\infty} \frac{(2a)^{2n_{2}+1}}{(2n_{2}+1)!} \, 
\Gamma( n_{2} + \tfrac{3}{4}) 
\Gamma( n_{2} + m + \tfrac{5}{4}). \nonumber 
\end{eqnarray}

Using the standard formulas (\ref{reduce-1}) and (\ref{reduce-2}), we can 
write this in the form 

\begin{eqnarray}
I_{1,3} &  =   & 
\frac{\Gamma(\tfrac{1}{4}) \Gamma( m + \tfrac{3}{4})}{4 \Gamma(m+1)} 
\, \, {_{2}F_{1}} 
\left(
\begin{matrix}
\tfrac{1}{4}  &  & m+ \tfrac{3}{4}   \\
 & \tfrac{1}{2} &   
\end{matrix}
\, \Big{|} a^{2} 
\right) - \nonumber \\
&  & \frac{a \Gamma(\tfrac{3}{4}) \Gamma( m + \tfrac{5}{4})}{2 \Gamma(m+1)} 
\, \, {_{2}F_{1}} 
\left(
\begin{matrix}
\tfrac{3}{4}  &  & m+ \tfrac{5}{4}   \\
 & \tfrac{3}{2} &   
\end{matrix}
\, \Big{|} a^{2} 
\right). 
\nonumber
\end{eqnarray}

In summary: we have obtained three series related to the integral 
$N_{0,4}(a,m)$. The series $I_{1,2}$ and $I_{2,3}$ are given in terms of 
the hypergeometric function ${_{2}F_{1}}$ with last argument $1/a^{2}$. 
These series converge when $a^{2} > 1$. The remaining case $I_{1,3}$ gives
${_{2}F_{1}}$ with argument $a^{2}$, that is convergent when $a^{2} < 1$.
Rule \ref{rule-disc} states that we must add the series $I_{1,2}$ and 
$I_{2,3}$ to get a valid representation for $a^{2} >1$. In conclusion, the 
method of brackets shows that
\begin{eqnarray}
N_{0,4}(a,m) &  =  & 
\frac{\Gamma(\tfrac{1}{4}) \Gamma(m + \tfrac{3}{4} ) }{4 \Gamma(m+1)} 
\, \, {_{2}F_{1}} 
\left(
\begin{matrix}
\tfrac{1}{4}  &  & m+ \tfrac{3}{4}   \\
 & \tfrac{1}{2} &   
\end{matrix}
\, \Big{|} a^{2} 
\right) + 
\nonumber \\
& -  & \frac{a \Gamma(\tfrac{3}{4}) \Gamma(m + \tfrac{5}{4} ) }{2 \Gamma(m+1)} 
\, \, {_{2}F_{1}} 
\left(
\begin{matrix}
\tfrac{3}{4}  &  & m+ \tfrac{5}{4}   \\
 & \tfrac{3}{2} &   
\end{matrix}
\, \Big{|} a^{2} 
\right) \quad \text{ for } a^{2} < 1, 
\nonumber \\
 &  =  & 
\frac{\Gamma(\tfrac{1}{2}) \Gamma(m + \tfrac{1}{2} ) }{2 \sqrt{2a} \Gamma(m+1)} 
\, \, {_{2}F_{1}} 
\left(
\begin{matrix}
\tfrac{1}{4}  &  & \tfrac{3}{4}   \\
 & \tfrac{1}{2} -m &   
\end{matrix}
\, \Big{|} \frac{1}{a^{2} }
\right) 
\nonumber \\
& + & \frac{ \Gamma(-\tfrac{1}{2}) \Gamma(2m + \tfrac{3}{2} ) }{2 
(2a)^{2m+3/2} \Gamma(m+1)} 
\, \, {_{2}F_{1}} 
\left(
\begin{matrix}
m+ \tfrac{3}{4}  &  & m+ \tfrac{5}{4}   \\
 & m+ \tfrac{3}{2} &   
\end{matrix}
\, \Big{|} \frac{1}{ a^{2} }
\right) \quad \text{ for } a^{2} > 1. 
\nonumber
\end{eqnarray}

The continuity of these expressions at $a=1$ requires the evaluation of 
${_{2}F_{1}}(a,b;c;1)$. Recall that this is finite only when $c > a+b$. In 
our case, we have four hypergeometric terms and in each one of them, the
corresponding expression $c-(a+b)$ equals $-\tfrac{1}{2}-m$. Therefore 
each hypergeometric term blows up as $a \to 1$. This divergence
is made evident by employing the relation
\begin{equation}
{_{2}F_{1}}(a,b,c;z) = (1-z)^{c-a-b} {_{2}F_{1}}(c-a, c-b,c;z).
\end{equation}
\noindent
The expression for $N_{0,4}(a,m)$ given above is transformed into 

\begin{eqnarray}
N_{0,4}(a,m) &  =  & 
\frac{\Gamma(\tfrac{1}{4}) \Gamma(m + \tfrac{3}{4} ) }{4 \Gamma(m+1) 
(1-a^{2})^{m+1/2}} 
\, \, {_{2}F_{1}} 
\left(
\begin{matrix}
\tfrac{1}{4}  &  & -m- \tfrac{1}{4}   \\
 & \tfrac{1}{2} &   
\end{matrix}
\, \Big{|} a^{2} 
\right) + 
\nonumber \\
& -  &  \frac{a \Gamma(\tfrac{3}{4}) \Gamma(m + \tfrac{5}{4} ) }
{2 \Gamma(m+1)(1-a^{2})^{m+1/2}} 
\, \, {_{2}F_{1}} 
\left(
\begin{matrix}
\tfrac{3}{4}  &  & -m+ \tfrac{1}{4}   \\
 & \tfrac{3}{2} &   
\end{matrix}
\, \Big{|} a^{2} 
\right) \quad \text{ for } a^{2} < 1, 
\nonumber \\
 &  =  & 
\frac{\Gamma(\tfrac{1}{2}) \Gamma(m + \tfrac{1}{2} ) }
{2 \sqrt{2a} \Gamma(m+1) (1-a^{-2})^{m+1/2}} 
\, \, {_{2}F_{1}} 
\left(
\begin{matrix}
\tfrac{1}{4}-m  &  & -\tfrac{1}{4} -m  \\
 & \tfrac{1}{2} -m &   
\end{matrix}
\, \Big{|} \frac{1}{a^{2} }
\right) 
\nonumber \\
& + & \frac{ \Gamma(-\tfrac{1}{2}) \Gamma(2m + \tfrac{3}{2} ) }{2 
(2a)^{2m+3/2} \Gamma(m+1) (1- a^{-2})^{m+1/2}} 
\, \, {_{2}F_{1}} 
\left(
\begin{matrix}
\tfrac{3}{4}  &  & \tfrac{1}{4}   \\
 & m+ \tfrac{3}{2} &   
\end{matrix}
\, \Big{|} \frac{1}{ a^{2} }
\right) \quad \text{ for } a^{2} > 1. 
\nonumber
\end{eqnarray}

\medskip

Introduce the functions
\begin{eqnarray}
G_{1}(a,m) & = & \small{\left( \frac{3}{4} \right)_{m} }
\, \, 
\small{
{_{2}F_{1}} 
\left(
\begin{matrix}
\tfrac{1}{4}  &  & -\tfrac{1}{4} - m   \\
 &  \tfrac{1}{2} &   
\end{matrix}
\, \Big{|} a^{2} 
\right) 
}
\nonumber \\
 & - & 2 a \small{ \left( \frac{1}{4} \right)_{m+1} }
\, \, 
\small{
{_{2}F_{1}} 
\left(
\begin{matrix}
\tfrac{3}{4}  &  & \tfrac{1}{4} - m   \\
 &  \tfrac{3}{2} &   
\end{matrix}
\, \Big{|} a^{2} 
\right),
}
\nonumber 
\end{eqnarray}
\noindent
and
\begin{eqnarray}
G_{2}(a,m) & = & \small{\left( \frac{1}{2} \right)_{m}}
 (2a)^{2m+1}
\, \, {_{2}F_{1}} 
\left(
\begin{matrix}
\tfrac{1}{4}-m  &  & -\tfrac{1}{4} - m   \\
 &  \tfrac{1}{2}-m &   
\end{matrix}
\, \Big{|} \frac{1}{a^{2}}
\right) \nonumber \\
 & - & (-1)^{m} m! 2^{-2m} 
\small{\binom{4m+1}{2m} }
\, \, 
\small{
{_{2}F_{1}} 
\left(
\begin{matrix}
\tfrac{3}{4}  &  & \tfrac{1}{4}  \\
 &  m+ \tfrac{3}{2} &   
\end{matrix}
\, \Big{|} \frac{1}{a^{2} }
\right). 
}
\nonumber 
\end{eqnarray}
\noindent
Then
\begin{equation}
N_{0,4}(a,m)  =  
\frac{\pi \sqrt{2} }{4m!} \frac{G_{1}(a,m)}{(1-a^{2})^{m+1/2}} 
\label{nzero-1}
\end{equation}
\noindent
for $a^{2} < 1$ and 
\begin{equation}
N_{0,4}(a,m)  =  \frac{\pi}{2^{2m+5/2} \sqrt{a} m!} 
\frac{G_{2}(a,m)}{(a^{2}-1)^{m+1/2}}  
\label{nzero-2}
\end{equation}
\noindent
for $a^{2} > 1$. The functions $G_{1}(a,m)$
and $G_{2}(a,m)$ match at $a=1$ to sufficiently high order to verify the
continuity at $a=1$. Morever, their blow up at $a=-1$ is a reflection of the 
fact that the convergence of the integral $N_{0,4}(a,m)$ 
requires $a> -1$. \\

It is possible to show that both expressions (\ref{nzero-1}) and 
(\ref{nzero-2})
reduce to (\ref{value-quartic}).
The details will appear elsewhere.

\section{Integrals from Feynman diagrams} \label{sec-feynman}
\setcounter{equation}{0}

The flexibility of the method of brackets is now illustrated by evaluating 
examples of definite integrals appearing in the resolution of Feynman diagrams. 
The reader will find in 
\cite{huang-1}, \cite{smirnov1}, \cite{itzykson1} and \cite{zinn-1} information 
about these diagrams. The mathematical theory behind Quantum Field Theory 
and in particular to the role of Feynman diagrams can be obtained from 
\cite{folland-1} and \cite{connes-marcolli}. 

The graph $G$ contains $N$ {\em propagators} or {\em internal lines}, $L$ 
{\em loops} associated to independent internal momenta ${\mathbf{Q}}:= 
\{ Q_{1}, \cdots, Q_{L} \}$, $E$ independent external momenta 
$\mathbf{P} := \{ P_{1}, \cdots, P_{E} \}$ (therefore the diagram has $E+1$
{\em external lines}). The momentum $P_{j}, Q_{j}$ belong to 
$\mathbb{R}^{4}$ and the space $\mathbb{R}^{4}$ 
is equipped with the Minkowski metric. Therefore, for $A, \, B 
\in \mathbb{R}^{4}$, we have
\begin{equation}
A^{2} := A_{0}^{2} - A_{1}^{2} - A_{2}^{2} - A_{3}^{2}, 
\end{equation}
\noindent
and 
\begin{equation}
A \cdot B := A_{0}B_{0} - A_{1}B_{1} - A_{2}B_{2} - A_{3}B_{3}.
\end{equation}
Finally, each propagator has a {\em mass} $m_{j} \geq 0$ 
associated to it, collected in the vector $\mathbf{m} = 
(m_{1}, \cdots, m_{N})$. 

The method of dimensional regularization 
(see \cite{ryder} for details) gives an
integral expression in the momentum space that represents the diagram in 
$D = 4 - 2 \epsilon$ dimensions. In Minkowski space the integral is given by
\begin{equation}
G = G( \mathbf{P}, \mathbf{m}) := 
\int \frac{d^{D}Q_{1}}{i \pi^{D/2}} \cdots 
\int \frac{d^{D}Q_{L}}{i \pi^{D/2}}  
\frac{1}{(B_{1}^{2}-m_{1}^{2})^{a_{1}}} 
\cdots 
\frac{1}{(B_{N}^{2}-m_{N}^{2})^{a_{N}}}. 
\label{feyn-1}
\end{equation}
\noindent
The symbol $B_{j}$ represents the momentum of the $j$-th propagator and it is
a linear combination of the internal and external momenta $\mathbf{P}$ and 
$\mathbf{Q}$, respectively. The vector $\mathbf{a} := ( a_{1}, \cdots, a_{N})$
captures the {\em powers} 
of the propagators and they may assume arbitrary values. 

In order to simplify (\ref{feyn-1}), we use the identity
\begin{equation}
\frac{1}{A^{\alpha}} = \frac{1}{\Gamma(\alpha)} \ift x^{\alpha-1} e^{-Ax} \, dx 
\end{equation}
\noindent
with $A = B_{j}^{2}-m_{j}^{2}$ and convert it into
\begin{equation}
G = \frac{1}{\prod_{j=1}^{N} \Gamma(a_{j})} 
\ift \text{exp} \left( \sum_{j=1}^{N} x_{j}m_{j}^{2} \right) 
\int \frac{\prod_{j=1}^{L} d^{D}Q_{j}}{(i \pi^{D/2})^{L} }
\text{exp} \left( - \sum_{j=1}^{N} x_{j} B_{j}^{2} \right) \mathbf{dx},
\label{feyn-2}
\end{equation}
\noindent
where $\displaystyle{\mathbf{dx} = \prod_{j=1}^{N} x_{j}^{a_{j}-1} dx_{j}}$. 

The next step in the reduction process is to integrate (\ref{feyn-2}) with 
respect to the internal momenta $Q_{j}$. This 
gives an expression for the integral 
$G$ in terms of only the external momenta $P_{j}$ and the masses $m_{j}$. This 
step can be
achieved by introducing the {\em Schwinger 
parametrizaton} (see \cite{gonzalez-2005} and chapter 1, section 4 of 
\cite{connes-marcolli}
for details) and $x_{j}$ are called the {\em Schwinger variables}. 
The final result is the representation
\begin{equation}
G = \frac{(-1)^{-LD/2} }{\prod_{j=1}^{N} \Gamma(a_{j}) } 
\ift U^{-D/2} \text{exp} \left( \sum_{j=1}^{N} x_{j}m_{j}^{2} \right) 
\text{exp} \left( - \frac{F}{U} \right) \mathbf{dx}. 
\label{Schwinger-rep}
\end{equation}
\noindent
The function $F$ corresponds to a quadratic structure of the external 
momentum defined by
\begin{equation}
F = \sum_{i,j=1}^{E} C_{i,j} P_{i} \cdot P_{j}.
\end{equation}
\noindent
The function $U$ and the coefficients $C_{i,j}$ are the 
{\em Symanzik polynomials} in the 
Schwinger parameters $x_{j}$. These polynomials are 
 given in terms of determinants of the so-called 
{\em matrix of parameters}. The polynomial $C_{i,j}$
are symmetric, that is $C_{i,j} = C_{j,i}$. A systematic algorithm to 
write down the expression (\ref{Schwinger-rep}) directly from the 
Feynman diagram is presented in \cite{gonzalez-2005}.

\begin{example}
Figure \ref{figure1}  depicts  the interaction of three 
particles corresponding to the three external lines of 
momentum $P_{1}, \, P_{2}, \, P_{3}$. In this case the Schwinger 
parametrization provides the integral
{{
\begin{figure}[ht]
\begin{center}
\centerline{\epsfig{file=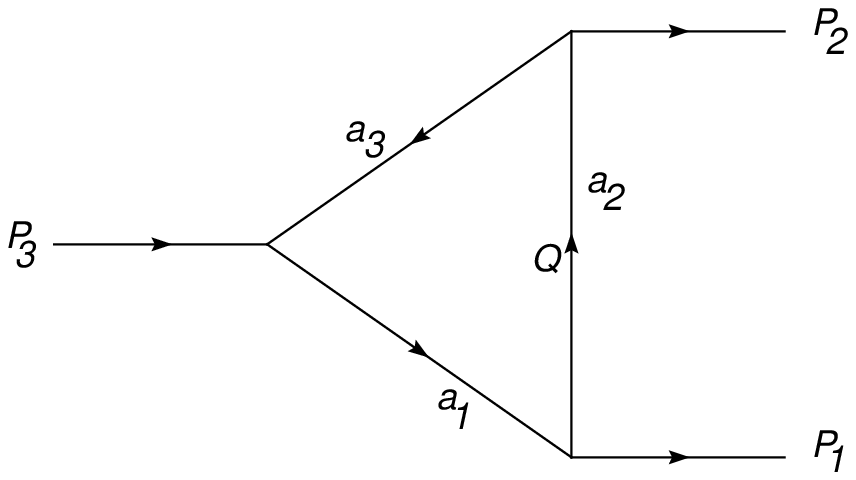,width=20em,angle=0}}
\caption{The triangle}
\label{figure1}
\end{center}
\end{figure}
}}

\begin{eqnarray}
G  & = & \frac{(-1)^{-D/2}}{\Gamma(a_{1})\Gamma(a_{2})\Gamma(a_{3})} 
\ift \ift \ift \frac{x_{1}^{a_{1}-1} x_{2}^{a_{2}-1} x_{3}^{a_{3}-1} }
{(x_{1}+x_{2}+x_{3})^{D/2}} 
\nonumber \\
& \times  & \text{exp}(x_{1}m_{1}^{2} + x_{2}m_{2}^{2} + x_{3}m_{3}^{2}) 
\text{exp} \left( - 
\frac{C_{1}P_{1}^{2} + 2C_{12}P_{1} \cdot P_{2} + C_{22}P_{2}^{2} }
{x_{1}+x_{2}+x_{3}} \right) dx_{1}dx_{2}dx_{3}. 
\nonumber
\end{eqnarray}
\noindent
The algorithm in \cite{gonzalez-2005} and \cite{gonzalez-2007} gives the coefficients $C_{i,j}$ as
\begin{equation}
C_{11} = x_{1}(x_{2}+x_{3}), \, \, C_{12} = x_{1}x_{3}, \, \, 
C_{22} = x_{3}(x_{1}+x_{2}). 
\end{equation}

Conservation of momentum gives $P_{3} = P_{1} + P_{2}$ and replacing the 
coefficients $C_{i,j}$ we obtain
\begin{eqnarray}
G & = & \frac{(-1)^{-D/2}}{\prod_{j=1}^{3} \Gamma(a_{j})} 
\ift \ift \ift x^{a_{1}-1} x^{a_{2}-1} x^{a_{3}-1} \times \nonumber \\
& \times & \frac{\text{exp}\left(x_{1}m_{1}^{2} + x_{2}m_{2}^{2} + 
x_{3}m_{3}^{2} \right)
\text{exp}\left( - \frac{x_{1}x_{2}P_{1}^{2} + x_{2}x_{3}P_{2}^{2} 
+ x_{3}x_{1}P_{3}^{2}}{x_{1}+x_{2}+x_{3}} \right)}{(x_{1}+x_{2}+x_{3})^{D/2}}
dx_{1}dx_{2}dx_{3}.
\nonumber
\end{eqnarray}

To solve the Feynman diagram in Figure \ref{figure1} 
it is required to evaluate the 
integral $G$ as a function of the variables $P_{1}, \, P_{2} \in 
{\mathbb{R}}^{4}$, the masses $m_{i}$, the dimension $D$ and the 
parameters $a_{i}$. 

We now describe the evaluation of the integral $G$ in the special massless 
situation: $m_{1}=m_{2}=m_{3} =0$. Moreover we assume that 
$P_{1}^{2}=P_{2}^{2}=0$. The 
integral to be evaluated is then
\begin{equation}
G_{1}  = \frac{(-1)^{-D/2}}{\Gamma(a_{1}) \Gamma(a_{2}) \Gamma(a_{3})}
\int_{\mathbb{R}_{+}^{3}} x_{1}^{a_{1}-1} x_{2}^{a_{2}-1} x_{3}^{a_{3}-1} 
\frac{\text{exp}\left( - \frac{x_{1}x_{3}}
{x_{1}+x_{2}+x_{3}} P_{3}^{2} \right)}{(x_{1}+x_{2}+x_{3})^{D/2}}
\, dx_{1}\, dx_{2}\, dx_{3}.
\nonumber
\end{equation}

The method of brackets gives 
\begin{equation}
G_{1} \eqf \frac{(-1)^{-D/2}}{\Gamma(a_{1}) \Gamma(a_{2}) \Gamma(a_{3}) }
\sum_{n_{1}} \sum_{n_{2}} \sum_{n_{3}} \sum_{n_{4}} \phi_{1234} 
(P_{3}^{2})^{n_{1}} \frac{\Delta_{1}\Delta_{2} \Delta_{3} \Delta_{4}}
{\Gamma( D/2+n_{1})},
\end{equation}
\noindent
where the brackets $\Delta_{j}$ are given by 
\begin{eqnarray}
\Delta_{1} & = & \langle{ D/2 + n_{1} + n_{2} + n_{3} + n_{4} \rangle}, 
\nonumber \\
\Delta_{2} & = & \langle{ a_{1} + n_{1} + n_{2} \rangle}, 
\nonumber \\
\Delta_{3} & = & \langle{ a_{2} + n_{3} \rangle}, 
\nonumber \\
\Delta_{4} & = & \langle{ a_{3} + n_{1} + n_{4} \rangle}. 
\nonumber 
\end{eqnarray}
\noindent
The solution contains no free indices: there are four sums and 
the linear system corresponding to the vanishing of the 
brackets eliminates all of them:
\begin{equation}
n_{1}^{*} = \tfrac{D}{2} - a_{1} - a_{2} - a_{3}, \, 
n_{2}^{*} = - \tfrac{D}{2} + a_{2} + a_{3}, \, 
n_{3}^{*} = -a_{2}, \, 
n_{4}^{*} = - \tfrac{D}{2} + a_{1} + a_{2}. \nonumber
\end{equation}
\noindent
We conclude that
\begin{eqnarray}
G_{1} &  = & \frac{(-1)^{-D/2}}{\Gamma(a_{1}) \Gamma(a_{2}) \Gamma(a_{3}) }
(P_{3}^{2})^{D/2-a_{1}-a_{2}-a_{3}} \times \nonumber \\
& \times & 
\frac{\Gamma(a_{1}+a_{2}+a_{3}-\tfrac{D}{2}) \Gamma( \tfrac{D}{2} -a_{2}-a_{3})
\Gamma(a_{2}) \Gamma(\tfrac{D}{2}) \Gamma(\tfrac{D}{2}-a_{1}-a_{2})}
{\Gamma(D - a_{1}-a_{2}-a_{3})}.
\nonumber 
\end{eqnarray}
\end{example}

\begin{example}
\label{example-bubble}
The second example considers the diagram depicted in Figure \ref{figure2}. The
resolution of this diagram is well-known and it appears in \cite{boos1},
\cite{davydychev1991} and \cite{davydychev1992}. The diagram contains 
two external lines and two internal lines (propagators) with the same mass $m$. 
These propagators are marked $1$ and $2$. 
{{
\begin{figure}[ht]
\begin{center}
\centerline{\epsfig{file=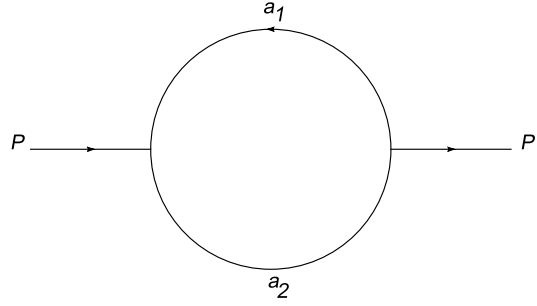,width=20em,angle=0}}
\caption{The bubble}
\label{figure2}
\end{center}
\end{figure}
}}

In momentum variables, the integral representation of this diagram is given 
by
\begin{equation}
G = \int_{\mathbb{R}^{D}} \frac{d^{D}Q}{i \pi^{D/2}} 
\frac{1}{(Q^{2}-m^{2})^{a_{1}} \, ( (P-Q)^{2} - m^{2})^{a_{2}}}. 
\end{equation}
\noindent
For the diagram considered here, we have $U = x_{1} + x_{2}$ and 
$F = x_{1}x_{2}P^{2}$. According to 
(\ref{Schwinger-rep}), the Schwinger representation is given by 
\begin{eqnarray}
G & = & \frac{(-1)^{-D/2}}{\Gamma(a_{1})\Gamma(a_{2})}  \nonumber \\
 & \times & \ift \ift \frac{x^{a_{1}-1} x^{a_{2}-1}}{(x_{1}+x_{2})^{D/2}} 
\text{exp}\left( m^{2}(x_{1}+x_{2}) \right) 
\text{exp} \left( - \frac{x_{1}x_{2}}{x_{1}+x_{2}} P^{2} \right) 
\, dx_{1} \, dx_{2}.
\nonumber
\end{eqnarray}

In order to generate the bracket series for $G$, we expand first 
the exponential function to obtain
\begin{equation}
G \eqf \frac{(-1)^{-D/2}}{\Gamma(a_{1}) \Gamma(a_{2})} 
\sum_{n_{1},n_{2}} \phi_{1,2} (P^{2})^{n_{1}} (-m^{2})^{n_{2}}  
\int_{R^{2}_{+}} \frac{x_{1}^{n_{1}} x_{2}^{n_{2}} \, dx_{1} \, dx_{2}}
{(x_{1}+x_{2})^{D/2+n_{1}-n_{2}}}. \label{G-part1}
\end{equation}

Expanding now the term 
\begin{equation}
\frac{1}{(x_{1}+x_{2})^{D/2+n_{1}-n_{2} } } \eqf 
\sum_{n_{3},n_{4}} \phi_{3,4} \frac{x_{1}^{n_{3}} x_{2}^{n_{4}} }
{\Gamma(D/2 + n_{1}-n_{2})} \Delta_{1},
\end{equation}
\noindent
with $\Delta_{1} = \langle{\tfrac{D}{2} + n_{1}-n_{2}+n_{3}+n_{4} \rangle}$,
and replacing in (\ref{G-part1}) yields
\begin{equation}
G \eqf \frac{(-1)^{-D/2}}{\Gamma(a_{1}) \Gamma(a_{2})} 
\sum_{n_{1},\cdots,n_{4}} \phi_{1,2,3,4} 
\frac{(P^{2})^{n_{1}} (-m^{2})^{n_{2}} }{\Gamma(\tfrac{D}{2}+n_{1}-n_{2})}
\Delta_{1} \Delta_{2}\Delta_{3},
\end{equation}
\noindent
where
\begin{eqnarray}
\Delta_{1} & = & \langle{\tfrac{D}{2} + n_{1}-n_{2}+n_{3}+n_{4} \rangle},
\nonumber \\
\Delta_{2} & = & \langle{ a_{1} + n_{1} + n_{3} \rangle}, \nonumber \\
\Delta_{3} & = & \langle{ a_{2} + n_{1} + n_{4} \rangle}. \nonumber 
\end{eqnarray}
The expression for $G$ contains $4$ indices and the vanishing of the brackets
allows us to express all of them in terms of a single index. We will denote 
by $G_{j}$ the expression for $G$ where the index $n_{j}$ is free.  \\

\noindent
{\bf The sum $G_{1}$}: in this case the solution of the corresponding linear
system is 
\begin{equation}
n_{2}^{*} = \tfrac{D}{2} - a_{1}-a_{2} - n_{1}, \, 
n_{3}^{*} = -a_{1} -n_{1}, \, n_{4}^{*} = -a_{2}-n_{1},
\end{equation}
\noindent
and the sum $G_{1}$ becomes
\begin{eqnarray}
G_{1} & = & (-1)^{-D/2} \frac{(-m^{2})^{D/2-a_{1}+a_{2} } } {\Gamma(a_{1}) 
\Gamma(a_{2}) } \times \nonumber \\
 & \times & \sum_{n_{1}=0}^{\infty} 
\frac{\Gamma(a_{1}+a_{2}- D/2 + n_{1}) \, \Gamma(a_{1}+n_{1}) \, 
\Gamma(a_{2}+n_{1}) }{\Gamma(a_{1}+a_{2}+2n_{1}) } 
\frac{ \left( \frac{P^{2}}{m^{2}} \right)^{n_{1}} }{n_{1}!}. \nonumber
\end{eqnarray}
\noindent
This can be expressed as 
\begin{equation}
G_{1} = \lambda_{1} (-m^{2})^{D/2-a_{1}+a_{2}} 
{_{3}F_{2}} 
\left(
\begin{matrix}
a_{1}+a_{2}- \tfrac{D}{2}, & a_{1}, & a_{2} \\
\tfrac{1}{2}(a_{1}+a_{2}+1), & \tfrac{1}{2}(a_{1}+a_{2}) &   
\end{matrix}
\Big{|} \frac{P^{2}}{4m^{2}}, 
\right)
\end{equation}
\noindent
where 
\begin{equation}
\lambda_{1} = (-1)^{-D/2} \frac{\Gamma(a_{1}+a_{2} - D/2)}{\Gamma(a_{1}+a_{2})}.
\end{equation}

\medskip

\noindent
{\bf The sum $G_{2}$}: keeping $n_{2}$ as the free index gives 
\begin{equation}
n_{1}^{*} = \tfrac{D}{2} - a_{1}-a_{2} - n_{2}, \, 
n_{3}^{*} = a_{2} -\tfrac{D}{2}+n_{2}, \, n_{4}^{*} = a_{1}- \tfrac{D}{2}
+n_{2},
\nonumber
\end{equation}
\noindent
which leads to 
\begin{equation}
G_{2} = \lambda_{2} (P_{1}^{2})^{D/2-a_{1}+a_{2}} 
{_{3}F_{2}} 
\left(
\begin{matrix}
a_{1}+a_{2}- \tfrac{D}{2}, & \tfrac{1}{2}(1+a_{1}+a_{2}-D), & 
\tfrac{1}{2}(2+a_{1}+a_{2}-D)  \\
1+a_{1}-\tfrac{D}{2}, & 1 + a_{2}- \tfrac{D}{2} &   
\end{matrix}
\Big{|} \frac{4m^{2}}{P^{2}} 
\right)
\nonumber
\end{equation}
\noindent
where the prefactor $\lambda_{2}$ is given by
\begin{equation}
\lambda_{2} = (-1)^{-D/2} \frac{\Gamma(a_{1}+a_{2} - D/2) 
\Gamma(\tfrac{D}{2}-a_{1})   \Gamma(\tfrac{D}{2} - a_{2}) }
{\Gamma(a_{1}) \Gamma(a_{2}) \Gamma(D - a_{1}-a_{2})}.
\nonumber
\end{equation}

\medskip

\noindent
{\bf The cases $G_{3}$ and $G_{4}$} are computed by a similar procedure.
The results are

\begin{equation}
G_{3} = \lambda_{3} (P_{1}^{2})^{-a_{1}} (-m^{2})^{D/2-a_{2}}
{_{3}F_{2}} 
\left(
\begin{matrix}
a_{1} , & \tfrac{1}{2}(1+a_{1}-a_{2}), & 
\tfrac{1}{2}(2+a_{1}-a_{2})  \\
1+a_{1}-a_{2}, & 1 -a_{2} + \tfrac{D}{2} &   
\end{matrix}
\Big{|} \frac{4m^{2}}{P^{2}} 
\right)
\nonumber
\end{equation}

\noindent
and 

\begin{equation}
G_{4} = \lambda_{4} (P_{1}^{2})^{-a_{2}} (-m^{2})^{D/2-a_{1}}
{_{3}F_{2}} 
\left(
\begin{matrix}
a_{2} , & \tfrac{1}{2}(1-a_{1}+a_{2}), & 
\tfrac{1}{2}(2-a_{1}+a_{2})  \\
1-a_{1}+a_{2}, & 1 -a_{1} + \tfrac{D}{2} &   
\end{matrix}
\Big{|} \frac{4m^{2}}{P^{2}} 
\right)
\nonumber
\end{equation}

\noindent
where the prefactors $\lambda_{3}$ and $\lambda_{4}$ are given by

\begin{equation}
\lambda_{3} = (-1)^{-D/2} \frac{\Gamma(a_{2} - D/2) }
{\Gamma(a_{2}) } \text{ and }
\lambda_{4} = (-1)^{-D/2} \frac{\Gamma(a_{1} - D/2) }
{\Gamma(a_{1}) }. 
\end{equation}

The contributions of these four sums are now classified according to their
region of convergence. This is determined by the parameter $\rho := 
|4m^{2}/P^{2}|$. In the region $\rho > 1$, only the sum $G_{1}$ converges, 
therefore $G = G_{1}$ there. In the region $\rho < 1$ the three remaining sums
converge. Therefore, according to Rule \ref{rule-disc}, we have 
\begin{equation}
G = \begin{cases}
     \begin{matrix}
     G_{1}  & \quad \text{ for }  & \rho > 1, \\
     G_{2} + G_{3} + G_{4}  & \quad \text{ for }  & \rho < 1. 
      \end{matrix}
\end{cases}
\end{equation}
We have evaluated the Feynman diagram in Figure \ref{figure2} and 
expressed its solution in terms of hypergeometric functions that correspond
naturally to the two quotient of the two energy scales present in the diagram. 
\end{example}
 
\medskip

\begin{example}
The next example shows that the method of brackets succeeds in the 
evaluation of very complicated integrals. 
We consider a Feynman diagram 
with four loops as shown in Figure \ref{figure3}. 

{{
\begin{figure}[ht]
\begin{center}
\centerline{\epsfig{file=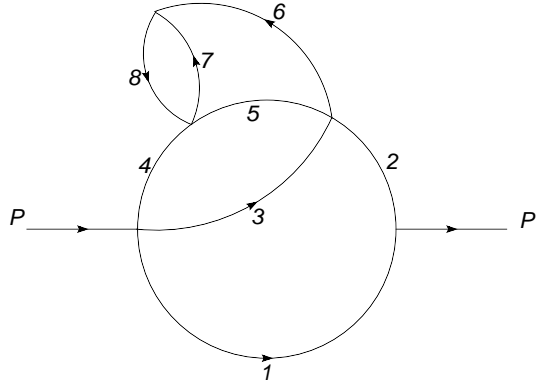,width=20em,angle=0}}
\caption{A diagram with four loops}
\label{figure3}
\end{center}
\end{figure}
}}

\medskip

The methods described in \cite{gonzalez-2005} for the 
Schwinger representation (\ref{Schwinger-rep}) of this diagram, give 
\begin{equation}
\begin{array}{ll}
U= &
x_{1}x_{3}x_{5}x_{7}+x_{1}x_{3}x_{5}x_{8}+x_{1}x_{3}x_{6}x_{7}+x_{1}x_{4}x_{5}x_{7}+x_{2}x_{3}x_{5}x_{7}+
\\
&
x_{1}x_{3}x_{6}x_{8}+x_{1}x_{4}x_{5}x_{8}+x_{1}x_{4}x_{6}x_{7}+x_{2}x_{3}x_{5}x_{8}+x_{2}x_{3}x_{6}x_{7}+
\\
&
x_{2}x_{4}x_{5}x_{7}+x_{1}x_{3}x_{7}x_{8}+x_{1}x_{4}x_{6}x_{8}+x_{1}x_{5}x_{6}x_{7}+x_{2}x_{3}x_{6}x_{8}+
\\
&
x_{2}x_{4}x_{5}x_{8}+x_{2}x_{4}x_{6}x_{7}+x_{3}x_{4}x_{5}x_{7}+x_{1}x_{4}x_{7}x_{8}+x_{1}x_{5}x_{6}x_{8}+
\\
&
x_{2}x_{3}x_{7}x_{8}+x_{2}x_{4}x_{6}x_{8}+x_{2}x_{5}x_{6}x_{7}+x_{3}x_{4}x_{5}x_{8}+x_{3}x_{4}x_{6}x_{7}+
\\
&
x_{1}x_{5}x_{7}x_{8}+x_{2}x_{4}x_{7}x_{8}+x_{2}x_{5}x_{6}x_{8}+x_{3}x_{4}x_{6}x_{8}+x_{3}x_{5}x_{6}x_{7}+
\\
&
x_{2}x_{5}x_{7}x_{8}+x_{3}x_{4}x_{7}x_{8}+x_{3}x_{5}x_{6}x_{8}+x_{3}x_{5}x_{7}x_{8}%
\end{array}%
\end{equation}%
\noindent
and for the function $F$ in (\ref{Schwinger-rep}): 

\begin{equation}
\begin{array}{ll}
F= &
(x_{1}x_{2}x_{3}x_{5}x_{7}+x_{1}x_{2}x_{3}x_{5}x_{8}+x_{1}x_{2}x_{3}x_{6}x_{7}+x_{1}x_{2}x_{4}x_{5}x_{7}+
\\
&
x_{1}x_{2}x_{3}x_{6}x_{8}+x_{1}x_{2}x_{4}x_{5}x_{8}+x_{1}x_{2}x_{4}x_{6}x_{7}+x_{1}x_{3}x_{4}x_{5}x_{7}+
\\
&
x_{1}x_{2}x_{3}x_{7}x_{8}+x_{1}x_{2}x_{4}x_{6}x_{8}+x_{1}x_{2}x_{5}x_{6}x_{7}+x_{1}x_{3}x_{4}x_{5}x_{8}+
\\
&
x_{1}x_{3}x_{4}x_{6}x_{7}+x_{1}x_{2}x_{4}x_{7}x_{8}+x_{1}x_{2}x_{5}x_{6}x_{8}+x_{1}x_{3}x_{4}x_{6}x_{8}+
\\
&
x_{1}x_{3}x_{5}x_{6}x_{7}+x_{1}x_{2}x_{5}x_{7}x_{8}+x_{1}x_{3}x_{4}x_{7}x_{8}+x_{1}x_{3}x_{5}x_{6}x_{8}+
\\
& x_{1}x_{3}x_{5}x_{7}x_{8})\;P^{2}.%
\end{array}%
\end{equation}%

The large number of terms appearing in the expressions for $U$ and $F$ ($34$ 
and $21$ respcetively) makes it almost impossible to apply the method of 
brackets without an apriori factorization of these polynomials.  This 
factorizations minimizes the number of sums and maximizes the number of 
brackets. In this example, the optimal factorization is given by

\begin{equation}
\begin{array}{l}
F=x_{1}f_{7}\;P^{2}, \\
\\
U=x_{1}f_{6}+f_{7},%
\end{array}%
\end{equation}%
where the functions $f_{i}$ are given by:

\begin{equation}
\begin{array}{l}
f_{7}=(x_{2}f_{6}+f_{5}), \\
f_{6}=x_{3}f_{4}+(x_{4}f_{4}+f_{3}), \\
f_{5}=x_{3}(x_{4}f_{4}+f_{3}), \\
f_{4}=x_{5}f_{2}+(x_{6}f_{2}+f_{1}), \\
f_{3}=x_{5}(x_{6}f_{2}+f_{1}), \\
f_{2}=(x_{7}+x_{8}), \\
f_{1}=x_{7}x_{8}.%
\end{array}%
\end{equation}

To analyze the diagram considered here, we start with the 
parametric representation

\begin{equation}
G=\dfrac{(-1)^{-D/2}}{\prod\limits_{j=1}^{8}\Gamma (a_{j})}
\int\limits_{0}^{\infty } \;\frac{\exp \left( -\dfrac{
x_{1}f_{7}}{x_{1}f_{6}+f_{7}}p_{1}^{2}\right) }{\left(
x_{1}f_{6}+f_{7}\right) ^{\frac{D}{2}}} {\mathbf{dx}},
\end{equation}
\noindent
and expand the exponential function first. A systematic expansion 
associated to the polynomials $f_{i}$ leads to the order  

\begin{equation}
U\longrightarrow f_{7}\longrightarrow f_{6}\longrightarrow
f_{5}\longrightarrow f_{4}\longrightarrow f_{3}\longrightarrow f_{2},
\end{equation}%

\noindent
that yields the bracket series

\begin{equation}
G  \eqf \dfrac{(-1)^{-D/2}}{\prod\limits_{j=1}^{8}\Gamma (a_{j})}%
\sum\limits_{n_{1},..,n_{15}}\phi _{n_{1},..,n_{15}}\;\dfrac{(P^{2})^{n_{1}}%
}{\Gamma (\frac{D}{2}+n_{1})}\;\Omega _{\left\{ n\right\}
}\prod\limits_{j=1}^{15}\Delta _{j},
\end{equation}%
where we have defined the factor 

\begin{equation}
\Omega _{\left\{ n\right\} }=\dfrac{1}{\Gamma (-n_{1}-n_{3})\Gamma
(-n_{2}-n_{4})\Gamma (-n_{5}-n_{7})\Gamma (-n_{6}-n_{8})\Gamma
(-n_{9}-n_{11})\Gamma (-n_{10}-n_{12})},  \nonumber
\end{equation}%
and the corresponding brackets by

\begin{equation}
\begin{array}{lll}
\Delta _{1}=\left\langle \frac{D}{2}+n_{1}+n_{2}+n_{3}\right\rangle , &  &
\Delta _{9}=\left\langle a_{2}+n_{4}\right\rangle , \\
\Delta _{2}=\left\langle -n_{1}-n_{3}+n_{4}+n_{5}\right\rangle , &  & \Delta
_{10}=\left\langle a_{3}+n_{5}+n_{6}\right\rangle , \\
\Delta _{3}=\left\langle -n_{2}-n_{4}+n_{6}+n_{7}\right\rangle , &  & \Delta
_{11}=\left\langle a_{4}+n_{8}\right\rangle , \\
\Delta _{4}=\left\langle -n_{5}-n_{7}+n_{8}+n_{9}\right\rangle , &  & \Delta
_{12}=\left\langle a_{5}+n_{9}+n_{10}\right\rangle , \\
\Delta _{5}=\left\langle -n_{6}-n_{8}+n_{10}+n_{11}\right\rangle , &  &
\Delta _{13}=\left\langle a_{6}+n_{12}\right\rangle , \\
\Delta _{6}=\left\langle -n_{9}-n_{11}+n_{12}+n_{13}\right\rangle , &  &
\Delta _{14}=\left\langle a_{7}+n_{13}+n_{14}\right\rangle , \\
\Delta _{7}=\left\langle -n_{10}-n_{12}+n_{14}+n_{15}\right\rangle , &  &
\Delta _{15}=\left\langle a_{8}+n_{13}+n_{15}\right\rangle , \\
\Delta _{8}=\left\langle a_{1}+n_{1}+n_{2}\right\rangle . &  &
\end{array}%
\end{equation}%
There is a unique way to evaluate the series: the numbers of indices is 
the same as the number of brackets. Solving the corresponding linear system 
leads to 

\begin{equation}
G=(-1)^{-D/2}\frac{(P^{2})^{n_{1}^{*}}}{\Gamma (D/2+n_{1}^{*})}
\;\Omega _{\left\{
n^{*}\right\} }\;\frac{\prod\limits_{j=1}^{15}\Gamma (-n_{j}^{*})}{%
\prod\limits_{j=1}^{8}\Gamma (a_{j})},
\end{equation}%
where the values $n_{i}^{*}$ are given by

\begin{equation}
\nonumber
\begin{array}{lll}
n_{1}^{*}=2D-a_{1}-a_{2}-a_{3}-a_{4}-a_{5}-a_{6}-a_{7}-a_{8}, &  &
n_{9}^{*}=D-a_{5}-a_{6}-a_{7}-a_{8}, \\
n_{2}^{*}=a_{2}+a_{3}+a_{4}+a_{5}+a_{6}+a_{7}+a_{8}-2D, &  &
n_{10}^{*}=a_{6}+a_{7}+a_{8}-D, \\
n_{3}^{*}=a_{1}-\frac{D}{2}, &  & n_{11}^{*}=a_{5}-\frac{D}{2}, \\
n_{4}^{*}=-a_{2}, &  & n_{12}^{*}=-a_{6}, \\
n_{5}^{*}=\frac{3D}{2}-a_{3}-a_{4}-a_{5}-a_{6}-a_{7}-a_{8}, &  & n_{13}^{*}=\frac{D}{%
2}-a_{7}-a_{8}, \\
n_{6}^{*}=a_{4}+a_{5}+a_{6}+a_{7}+a_{8}-\frac{3D}{2}, &  & n_{14}^{*}=a_{8}-\frac{D}{%
2}, \\
n_{7}^{*}=a_{3}-\frac{D}{2}, &  & n_{15}^{*}=a_{7}-\frac{D}{2}, \\
n_{8}^{*}=-a_{4}. &  &
\end{array}%
\end{equation}
\end{example}

\medskip

\begin{example}
The last example discussed in this paper gives the value of a Feynman 
diagram as a hypergeometric function of two variables.  The diagram 
shown in Figure \ref{figure2} contains
two external lines and with internal lines (propagators) with distinct 
masses. The same diagram with equal masses was described in  Example 
\ref{example-bubble}. The integral 
representation of this diagram in the momentum space is given by
\begin{equation}
G=\int \frac{d^{D}Q}{i\pi ^{D/2}}\frac{1}{(Q^{2}-m_{1}^{2})^{a_{1}}\left(
(P-Q)^{2}-m_{2}^{2}\right) ^{a_{2}}}.
\end{equation}%
\noindent
On the other hand, the parametric representation of Schwinger is
\begin{equation}
G=\dfrac{(-1)^{-\frac{D}{2}}}{\prod\limits_{j=1}^{2}\Gamma (a_{j})}%
\int\limits_{0}^{\infty } \;\frac{\exp \left(
x_{1}m_{1}^{2}\right) \exp \left( x_{2}m_{2}^{2}\right) \exp \left( -\frac{%
x_{1}x_{2}}{x_{1}+x_{2}}P^{2}\right) }{\left( x_{1}+x_{2}\right) ^{\frac{D}{2%
}}} \, {\bf dx}.
\end{equation}%

In order to find the bracket series associated to this integral, we first 
expand the exponentials

\begin{equation}
G \eqf \dfrac{(-1)^{-\frac{D}{2}}}{\prod\limits_{j=1}^{2}\Gamma (a_{j})}%
\sum\limits_{n_{1},n_{2},n_{3}}\phi _{n_{1,}n_{2},n_{3}}\;\left(
-m_{1}^{2}\right) ^{n_{1}}\left( -m_{2}^{2}\right) ^{n_{2}}\left(
P^{2}\right) ^{n_{3}}\int \;\frac{%
x_{1}^{n_{1}+n_{3}}x_{2}^{n_{2}+n_{3}}}{\left( x_{1}+x_{2}\right) ^{\frac{D}{%
2}+n_{3}}} \, {\mathbf{dx}} .  \label{k}
\nonumber
\end{equation}%
\noindent
and then the denominator
\begin{equation}
\nonumber
\frac{1}{\left( x_{1}+x_{2}\right) ^{\frac{D}{2}+n_{3}}} \eqf
\sum\limits_{n_{4},n_{5}}\phi _{n_{4,}n_{5}}\;\frac{%
x_{1}^{n_{4}}x_{2}^{n_{5}}}{\Gamma (\frac{D}{2}+n_{3})}\left\langle \tfrac{D%
}{2}+n_{3}+n_{4}+n_{5}\right\rangle.
\end{equation}%
\noindent
We obtain
\begin{eqnarray}
\nonumber
G & \eqf & \frac{(-1)^{-\frac{D}{2}}}{\prod\limits_{j=1}^{2}\Gamma (a_{j})}%
\sum\limits_{n_{1},..,n_{5}}\phi _{n_{1,..,}n_{5}}\;\frac{\left(
-m_{1}^{2}\right) ^{n_{1}}\left( -m_{2}^{2}\right) ^{n_{2}}\left(
P^{2}\right) ^{n_{3}}}{\Gamma (\frac{D}{2}+n_{3})}\Delta _{1} \nonumber \\
& \times  & 
\int x_{1}^{a_{1}+n_{1}+n_{3}+n_{4}-1} \, dx_{1} 
\, \, \int x_{2}^{a_{2}+n_{2}+n_{3}+n_{5}-1} \, dx_{2}.
\nonumber
\end{eqnarray}%
\noindent
Using the rules for transforming integrals into brackets yields
\begin{equation}
G \eqf  \frac{(-1)^{-\frac{D}{2}}}{\prod\limits_{j=1}^{2}\Gamma
(a_{j})}\sum\limits_{n_{1},..,n_{5}}\phi _{n_{1,..,}n_{5}} \frac{\left(
-m_{1}^{2}\right) ^{n_{1}}\left( -m_{2}^{2}\right) ^{n_{2}}\left(
P^{2}\right) ^{n_{3}}}{\Gamma (\frac{D}{2}+n_{3})}
\Delta _{1} \Delta_{2} \Delta_{3},
\end{equation}%
\noindent
with brackets defined by
\begin{equation}
\begin{array}{l}
\Delta _{1}=\left\langle \frac{D}{2}+n_{3}+n_{4}+n_{5}\right\rangle, \\
\\
\Delta _{2}=\left\langle a_{1}+n_{1}+n_{3}+n_{4}\right\rangle, \\
\\
\Delta _{3}=\left\langle a_{2}+n_{2}+n_{3}+n_{5}\right\rangle.%
\end{array}%
\label{f34}
\end{equation}%
\noindent
We have to choose two free values from $n_{1}, \cdots, n_{5}$. The result
is a hypergeometric function of multiplicity two. There are 
$10$ such choices and the brackets in (\ref{f34}) produce the linear
system
\begin{equation}
\begin{array}{l}
0=\frac{D}{2}+n_{3}+n_{4}+n_{5} \\
\\
0=a_{1}+n_{1}+n_{3}+n_{4} \\
\\
0=a_{2}+n_{2}+n_{3}+n_{5}.%
\end{array}%
\end{equation}

We denote by $G_{i,j}$ the solution of this system 
with free indices $n_{i}$ and $n_{j}$. The
series appearing in the solution to the Feynman diagram 
in Figure \ref{figure3} is expressed in terms
of the Appell function $F_{4}$, defined by
\begin{equation}
 F_{4}\left( \left.
\begin{array}{ccc}
\alpha &  & \beta  \\
&  &  \\
\gamma  &  & \delta %
\end{array}
\right\vert \, x, \, y \right)
=\sum\limits_{m,n=0}^{\infty }\frac{%
(\alpha )_{m+n}(\beta )_{m+n}}{(\gamma )_{m}(\delta )_{n}}\frac{x^{m}}{m!}%
\dfrac{y^{n}}{n!}.
\end{equation}

Following a procedure similar to the one described in the previous example, 
we obtain 
the explicit values of the integral $G$ is given in terms of the functions 
$G_{i,j}$. 

\begin{equation}
G = 
\begin{cases}
\begin{matrix}
G_{1,2}+G_{1,4}+G_{2,5} & \text{ when } &
m_{1}^{2},m_{2}^{2}<P^{2}  \\
G_{1,3}+G_{3,5} & \text{ when }  & 
m_{1}^{2},P^{2}<m_{2}^{2} \\
G_{2,3}+G_{3,4}
& \text{ when } &m_{2}^{2},P^{2}<m_{1}^{2}.
\end{matrix}
\end{cases}
\nonumber 
\end{equation}

These in turn are expressed in terms of the Appell function: \\

{\small{
\begin{eqnarray}
\nonumber
G_{1,2} & = & (-1)^{-\frac{D}{2}}\frac{\Gamma (a_{1}+a_{2}-\frac{D}{2})\Gamma (%
\frac{D}{2}-a_{1})\Gamma (\frac{D}{2}-a_{2})}{\Gamma (a_{1})\Gamma
(a_{2})\Gamma (D-a_{1}-a_{2})}\left( P^{2}\right) ^{\frac{D}{2}-a_{1}-a_{2}}
\\
&  & \nonumber \\
& \times & F_{4}\left( \left.
\begin{array}{ccc}
1+a_{1}+a_{2}-D &  & a_{1}+a_{2}-\frac{D}{2} \\
&  &  \\
1+a_{1}-\frac{D}{2} &  & 1+a_{2}-\frac{D}{2}%
\end{array}
\right\vert \frac{m_{1}^{2}}{P^{2}},\frac{m_{2}^{2}}{P^{2}}\right)
\nonumber \\
& & \nonumber \\
\nonumber
G_{1,4} & = & (-1)^{-\frac{D}{2}}\frac{\Gamma (a_{2}-\frac{D}{2})}{\Gamma (a_{2})}%
\left( -m_{2}^{2}\right) ^{\frac{D}{2}-a_{2}}\left( P^{2}\right)
^{-a_{1}} \\ 
& \times & \;F_{4}\left( \left.
\begin{array}{ccc}
1+a_{1}-\frac{D}{2} &  & a_{1} \\
&  &  \\
1+a_{1}-\frac{D}{2} &  & 1-a_{2}+\frac{D}{2}%
\end{array}%
\right\vert \frac{m_{1}^{2}}{P^{2}},\frac{m_{2}^{2}}{P^{2}}\right)
\nonumber \\
& & \nonumber \\
\nonumber
G_{2,5} & = & (-1)^{-\frac{D}{2}}\frac{\Gamma (a_{1}-\frac{D}{2})}{\Gamma (a_{1})}%
\left( -m_{1}^{2}\right) ^{\frac{D}{2}-a_{1}}\left( P^{2}\right)
^{-a_{2}} \\
& \times & \;F_{4}\left( \left.
\begin{array}{ccc}
1+a_{2}-\frac{D}{2} &  & a_{2} \\
&  &  \\
1-a_{1}+\frac{D}{2} &  & 1+a_{2}-\frac{D}{2}%
\end{array}%
\right\vert \frac{m_{1}^{2}}{P^{2}},\frac{m_{2}^{2}}{P^{2}}\right)
\nonumber \\
& & \nonumber \\
\nonumber
G_{1,3} & = & (-1)^{-\frac{D}{2}}\frac{\Gamma (a_{1}+a_{2}-\frac{D}{2})\Gamma (%
\frac{D}{2}-a_{1})}{\Gamma (a_{2})\Gamma (\frac{D}{2})}\left(
-m_{2}^{2}\right) ^{\frac{D}{2}-a_{1}-a_{2}} \\
& \times & \;F_{4}\left( \left.
\begin{array}{ccc}
a_{1}+a_{2}-\frac{D}{2} &  & a_{1} \\
&  &  \\
\frac{D}{2} &  & 1+a_{1}-\frac{D}{2}%
\end{array}%
\right\vert \frac{P^{2}}{m_{2}^{2}},\frac{m_{1}^{2}}{m_{2}^{2}}\right)
\nonumber \\
& & \nonumber \\
\nonumber
G_{3,5} & = & (-1)^{-\frac{D}{2}}\frac{\Gamma (a_{1}-\frac{D}{2})}{\Gamma (a_{1})}%
\left( -m_{1}^{2}\right) ^{\frac{D}{2}-a_{1}}\left( -m_{2}^{2}\right)
^{-a_{2}} \\
& \times & \;F_{4}\left( \left.
\begin{array}{ccc}
\frac{D}{2} &  & a_{2} \\
&  &  \\
\frac{D}{2} &  & 1-a_{1}+\frac{D}{2}%
\end{array}%
\right\vert \frac{P^{2}}{m_{2}^{2}},\frac{m_{1}^{2}}{m_{2}^{2}}\right)
\nonumber \\
& & \nonumber \\
\nonumber
G_{2,3} & = & (-1)^{-\frac{D}{2}}\frac{\Gamma (a_{1}+a_{2}-\frac{D}{2})\Gamma (%
\frac{D}{2}-a_{2})}{\Gamma (a_{1})\Gamma (\frac{D}{2})}\left(
-m_{1}^{2}\right) ^{\frac{D}{2}-a_{1}-a_{2}} \\ 
& \times & 
\;F_{4}\left( \left.
\begin{array}{ccc}
a_{1}+a_{2}-\frac{D}{2} &  & a_{2} \\
&  &  \\
\frac{D}{2} &  & 1+a_{2}-\frac{D}{2}%
\end{array}%
\right\vert \frac{P^{2}}{m_{1}^{2}},\frac{m_{2}^{2}}{m_{1}^{2}}\right)
\nonumber \\
 & & \nonumber \\
\nonumber
G_{3,4} & = & 
(-1)^{-\frac{D}{2}}\frac{\Gamma (a_{2}-\frac{D}{2})}{\Gamma (a_{2})}%
\left( -m_{2}^{2}\right) ^{\frac{D}{2}-a_{2}}\left( -m_{1}^{2}\right)
^{-a_{1}} \\ 
& \times & \;F_{4}\left( \left.
\begin{array}{ccc}
\frac{D}{2} &  & a_{1} \\
&  &  \\
\frac{D}{2} &  & 1-a_{2}+\frac{D}{2}%
\end{array}%
\right\vert \frac{P^{2}}{m_{1}^{2}},\frac{m_{2}^{2}}{m_{1}^{2}}\right).
\nonumber
\end{eqnarray}
}}

\end{example}

\section{Conclusions and future work} \label{sec-conclusions}
\setcounter{equation}{0}

The method of brackets provides a very effective procedure to evaluate 
definite integrals over the interval $[0, \infty)$. The method is based on 
a heuristic list of rules on the bracket series associated to such 
integrals. In particular we have provided a variety of examples that 
illustrate the power of this method. A rigorous validation of these rules
as well as a systematic study of integrals from 
Feynman diagrams is in progress. \\

\noindent
{\bf Acknowledgments}. The authors wish to thank R. Crandall for discussions
on an earlier version of the paper. \\


\begin{thebibliography}{10}

\bibitem{adams1}
C.~Adams, M.~{H}ildebrand, and {J}. Weeks.
\newblock Hyperbolic invariants of knots and ideals.
\newblock {\em Trans. {A}mer. {M}ath. {S}oc.}, 326:1--56, 1991.

\bibitem{moll-gr5}
T.~Amdeberhan, L.~Medina, and V.~Moll.
\newblock The integrals in {G}radshteyn and {R}yzhik. {P}art 5: {S}ome
  trigonometric integrals.
\newblock {\em Scientia}, 15:47--60, 2007.

\bibitem{moll-gr7}
T.~Amdeberhan and V.~Moll.
\newblock The integrals in {G}radshteyn and {R}yzhik. {P}art 7: {E}lementary
  examples.
\newblock {\em Scientia}, 16:25--40, 2008.

\bibitem{amram}
T.~Amdeberhan and V.~Moll.
\newblock A formula for a quartic integral: a survey of old proofs and some new
  ones.
\newblock {\em Ramanujan Journal}, 2009.

\bibitem{anastasiou-a}
C.~Anastasiou, E.~W.~N. Glover, and C.~Oleari.
\newblock Application of the negative-dimension approach to massless scalar box
  integrals.
\newblock {\em Nucl. {P}hys. B}, 565:445--467, 2000.

\bibitem{anastasiou-b}
C.~Anastasiou, E.~W.~N. Glover, and C.~Oleari.
\newblock Scalar one-loop integrals using the negative-dimension approach.
\newblock {\em Nucl. {P}hys. B}, 572:307--360, 2000.

\bibitem{antimirov1}
M.~Ya. Antimirov, A.~A. Kolyshkin, and R.~Vaillancourt.
\newblock {\em Complex {V}ariables}.
\newblock Academic Press, 1998.

\bibitem{bierens1}
D.~Bierens~de Haan.
\newblock {\em Tables d'integrales definies}.
\newblock C. G. Van der Post, Amsterdam, 1st edition, 1858.

\bibitem{bierens2}
D.~Bierens~de Haan.
\newblock {\em Expose de la theorie, des proprietes, des formules de
  transformation, et des methodes d'evaluation des integrales definies}.
\newblock C. G. Van der Post, Amsterdam, 1st edition, 1862.

\bibitem{bierens3}
D.~Bierens~de Haan.
\newblock {\em Nouvelles tables d'integrales definies}.
\newblock P. Engels, Leiden, 1st edition, 1867.

\bibitem{bollini}
C.~G. Bollini and J.~J. Giambiagi.
\newblock Dimensional renormalization: the number of dimensions as a
  regularizing parameter.
\newblock {\em Nuovo {C}imento}, B12:20--25, 1972.

\bibitem{boos1}
E.~E. Boos and A.~I. Davydychev.
\newblock A method of evaluating massive {F}eynman integrals.
\newblock {\em Theor. Math. Phys.}, 89:1052--1063, 1991.

\bibitem{irrbook}
G.~Boros and V.~Moll.
\newblock {\em Irresistible {I}ntegrals}.
\newblock Cambridge {U}niversity {P}ress, {N}ew {Y}ork, 1st edition, 2004.

\bibitem{sarah1}
G.~Boros, V.~Moll, and S.~Riley.
\newblock An elementary evaluation of a quartic integral.
\newblock {\em Scientia}, 11:1--12, 2005.

\bibitem{bor-broad1998}
J.~M. Borwein and D.~J. Broadhurst.
\newblock Determination of rational {D}edekind zeta invariants of hyperbolic
  manifolds and {F}eynman knots and links. {\em {u}npublished manuscript
  available at} arxiv:hep-th/9811173 v1.

\bibitem{bronstein2}
M.~Bronstein.
\newblock {\em Symbolic Integration I. {T}ranscendental functions}, volume~1 of
  {\em Algorithms and Computation in Mathematics}.
\newblock Springer-Verlag, 1997.

\bibitem{cherry1}
G.~W. Cherry.
\newblock Integration in finite terms with special functions: the error
  function.
\newblock {\em J. Symb. Comput.}, 1:283--302, 1985.

\bibitem{cherry2}
G.~W. Cherry.
\newblock Integration in finite terms with special functions: the logarithmic
  function.
\newblock {\em SIAM J. Comput.}, 15:1--21, 1986.

\bibitem{cherry5}
G.~W. Cherry.
\newblock An analysis of the rational exponential integral.
\newblock {\em SIAM J. Comput.}, 18:893--905, 1989.

\bibitem{coffey2008}
M.~Coffey.
\newblock Evaluation of a $\ln \tan$ integral arising in quantum field theory.
\newblock {\em J. Math. Phys.}, 52:093508, 2008.

\bibitem{coffey2009}
M.~Coffey.
\newblock Alternative evaluation of a $\ln \tan$ integral arising in quantum
  field theory.
\newblock {\em arXiv:0810.5077v2 [math-ph]}, 2009.

\bibitem{connes-marcolli}
M.~Connes and M.~Marcolli.
\newblock {\em Noncommutative {G}eometry, {Q}uantum {F}ields and {M}otives},
  volume~55 of {\em Colloquium {P}ublications}.
\newblock American {M}athematical {S}ociety, 2007.

\bibitem{crandall-08}
R.~Crandall.
\newblock Personal communication.
\newblock 2008.

\bibitem{davydychev1991}
A.~I. Davydychev.
\newblock Some exact results for $n$-point massive {F}eynman integrals.
\newblock {\em Jour. Math. Phys.}, 32:1052--1060, 1991.

\bibitem{davydychev1992}
A.~I. Davydychev.
\newblock General results for massive $n$-point {F}eynman diagrams with
  different masses.
\newblock {\em Jour. Math. Phys.}, 33:358--369, 1992.

\bibitem{dunne-1987}
G.~V. Dunne and I.~G. Halliday.
\newblock Negative dimensional integration. 2. {P}ath integrals and fermionic
  equivalence.
\newblock {\em Phys. Lett. B}, 193:247, 1987.

\bibitem{dunne-1989}
G.~V. Dunne and I.~G. Halliday.
\newblock Negative dimensional oscillators.
\newblock {\em Nuclear Physics B}, 308:589--618, 1989.

\bibitem{edwards2}
J.~Edwards.
\newblock {\em A treatise on the {I}ntegral {C}alculus}.
\newblock MacMillan, New York, 1922.

\bibitem{espmoll1}
O.~Espinosa and V.~Moll.
\newblock On some definite integrals involving the {H}urwitz zeta function.
  {P}art 1.
\newblock {\em The {R}amanujan {J}ournal}, 6:159--188, 2002.

\bibitem{espmoll3}
O.~Espinosa and V.~Moll.
\newblock The evaluation of {T}ornheim double sums. {P}art 1.
\newblock {\em Journal of {N}umber {T}heory}, 116:200--229, 2006.

\bibitem{espmoll6}
O.~Espinosa and V.~Moll.
\newblock The evaluation of {T}ornheim double sums. {P}art 2.
\newblock {\em Submitted for publication}, 2008.

\bibitem{fichtenholz1}
G.~M. Fichtenholz.
\newblock {\em Course in Differential and Integral Calculus}, volume 1,2,3.
\newblock Moscow, 1948.

\bibitem{flanders1}
H.~Flanders.
\newblock On the {F}resnel integrals.
\newblock {\em Amer. {M}ath. {M}onthly}, 89:264--266, 1982.

\bibitem{folland-1}
G.~Folland.
\newblock {\em Quantum {F}ield {T}heory. {A} {T}ourist {G}uide for
  {M}athematicians}, volume 149 of {\em Colloquium {P}ublications}.
\newblock American {M}athematical {S}ociety, 2008.

\bibitem{gonzalez-2005}
I.~Gonzalez and I.~Schmidt.
\newblock Recursive method to obtain the parametric representation of a generic
  {F}eynman diagram.
\newblock {\em Phys. Rev. D}, 72:106006, 2005.

\bibitem{gonzalez-2007}
I.~Gonzalez and I.~Schmidt.
\newblock Optimized negative dimensional integration method ({NDIM}) and
  multiloop {F}eynman diagram calculation.
\newblock {\em Nuclear Physics B}, 769:124--173, 2007.

\bibitem{gonzalez-2008}
I.~Gonzalez and I.~Schmidt.
\newblock Modular application of an integration by fractional expansion
  ({IBFE}) method to multiloop {F}eynman diagrams.
\newblock {\em Phys. Rev. D}, 78:086003, 2008.

\bibitem{gr6}
I.~S. Gradshteyn and I.~M. Ryzhik.
\newblock {\em Table of {I}ntegrals, {S}eries, and {P}roducts}.
\newblock Edited by A. Jeffrey and D. Zwillinger. Academic Press, New York, 6th
  edition, 2000.

\bibitem{gr}
I.~S. Gradshteyn and I.~M. Ryzhik.
\newblock {\em Table of {I}ntegrals, {S}eries, and {P}roducts}.
\newblock Edited by A. Jeffrey and D. Zwillinger. Academic Press, New York, 7th
  edition, 2007.

\bibitem{halliday-1987}
I.~G. Halliday and R.~M. Ricotta.
\newblock Negative dimensional integrals. {I}. {F}eynman graphs.
\newblock {\em Phys. Lett. B}, 193:241, 1987.

\bibitem{huang-1}
K.~Huang.
\newblock {\em Quantum {F}ield {T}heory. {F}rom operators to path integrals}.
\newblock John Wiley and Sons Inc., 1st edition, 1998.

\bibitem{itzykson1}
C.~Itzykson and J.~B. Zuber.
\newblock {\em Quatum {F}ield {T}heory}.
\newblock Mc{G}raw-{H}ill {I}nternational {B}ook {C}o., 1st edition, 1980.

\bibitem{koutschan1}
C.~Koutschan and V.~Levandovskyy.
\newblock Computing one of {V}ictor {M}oll's irresistible integrals with
  computer algebra.
\newblock {\em Computer {S}cience {J}ournal of {M}oldova}, 16:35--49, 2008.

\bibitem{kreimer1}
D.~Kreimer.
\newblock {\em Knots and Feynman diagrams}.
\newblock Number~13 in Cambridge Lectures in Physics. Cambridge University
  Press, 2000.

\bibitem{leonard1}
I.~E. Leonard.
\newblock More on {F}resnel integrals.
\newblock {\em Amer. {M}ath. {M}onthly}, 95:431--433, 1988.

\bibitem{lewin1}
L.~Lewin.
\newblock {\em Dilogarithms and {A}ssociated {F}unctions}.
\newblock Elsevier, North Holland, 2nd. edition, 1981.

\bibitem{lindman1}
C.~F. Lindman.
\newblock {\em Examen des nouvelles tables d'integrales definies de M. Bierens
  de Haan}.
\newblock P.A. Norstedt and Soner, Stockholm, 1891.

\bibitem{lunev1994}
F.~A. Lunev.
\newblock Evaluation of two-loop self-energy diagram with three propagators.
\newblock {\em Phys. Rev. D}, 50:7735--7737, 1994.

\bibitem{maclachlan-reid}
C.~Maclachlan and A.~Reid.
\newblock {\em The arithmetic of hyperbolic $3$-manifolds}.
\newblock Springer-Verlag, New York, 2003.

\bibitem{manna-moll-survey}
D.~Manna and V.~Moll.
\newblock A remarkable sequence of integers.
\newblock Preprint, 2009.

\bibitem{prudnikov1}
A.~P. Prudnikov Yu. A. Brychkov O.~I. Marichev.
\newblock {\em Integrals and Series}.
\newblock Gordon and Breach Science Publishers, 1992.

\bibitem{luis2}
L.~Medina and V.~Moll.
\newblock A class of logarithmic integrals.
\newblock {\em Ramanujan Journal}, To appear, 2009.

\bibitem{menasco}
W.~Menasco and M.~Thistlewaite, editors.
\newblock {\em Handbook of {K}not {T}heory}.
\newblock Elsevier, 2005.

\bibitem{milnor82}
J.~Milnor.
\newblock Hyperbolic {G}eometry: the first $150$ years.
\newblock {\em Bull. {A}mer. {M}ath. {Soc.}}, 6:9--24, 1982.

\bibitem{moll-gr1}
V.~Moll.
\newblock The integrals in {G}radshteyn and {R}yzhik. {P}art 1: {A} family of
  logarithmic integrals.
\newblock {\em Scientia}, 14:1--6, 2007.

\bibitem{moll-gr2}
V.~Moll.
\newblock The integrals in {G}radshteyn and {R}yzhik. {P}art 2: {E}lementary
  logarithmic integrals.
\newblock {\em Scientia}, 14:7--15, 2007.

\bibitem{moll-gr3}
V.~Moll.
\newblock The integrals in {G}radshteyn and {R}yzhik. {P}art 3: {C}ombinations
  of logarithms and exponentials.
\newblock {\em Scientia}, 15:31--36, 2007.

\bibitem{moll-gr4}
V.~Moll.
\newblock The integrals in {G}radshteyn and {R}yzhik. {P}art 4: {T}he gamma
  function.
\newblock {\em Scientia}, 15:37--46, 2007.

\bibitem{moll-gr6}
V.~Moll.
\newblock The integrals in {G}radshteyn and {R}yzhik. {P}art 6: {T}he beta
  function.
\newblock {\em Scientia}, 16:9--24, 2008.

\bibitem{moll-gr8}
V.~Moll, J.~Rosenberg, A.~Straub, and P.~Whitworth.
\newblock The integrals in {G}radshteyn and {R}yzhik. {P}art 8: {C}ombinations
  of powers, exponentials and logarithms.
\newblock {\em Scientia}, 16:41--50, 2008.

\bibitem{aequalsb}
M.~Petkovsek, H.~Wilf, and D.~Zeilberger.
\newblock {\em A=B}.
\newblock A. K. Peters, Ltd., 1st edition, 1996.

\bibitem{ryder}
L.~H. Ryder.
\newblock {\em Quantum {F}ield {T}heory}.
\newblock Cambridge {U}niversity {P}ress, 2nd edition, 1996.

\bibitem{smirnov1}
V.~A. Smirnov.
\newblock {\em Feynman {I}ntegral {C}alculus}.
\newblock Springer Verlag, Berlin Heildelberg, 2006.

\bibitem{suzuki-int2}
A.~T. Suzuki.
\newblock Evaluating residues and integrals through {N}egative {D}imensional
  {I}ntegration {M}ethod ({NDIM}).
\newblock {\em math-ph/0407032}, 2004.

\bibitem{suzuki-int1}
A.~T. Suzuki.
\newblock Negative dimensional approach to evaluating real integrals.
\newblock {\em math-ph/0806.3216}, 2008.

\bibitem{suzuki-massive2}
A.~T. Suzuki, E.~S. Santos, and A.~G.~M. Schmidt.
\newblock General massive one-loop off-shell three-point functions.
\newblock {\em Jour. Phys. A}, 36:4465--4476, 2003.

\bibitem{suzuki-massive1}
A.~T. Suzuki and A.~G.~M. Schmidt.
\newblock Negative dimensional integration for massive four point functions.
  {II}. new solutions.
\newblock {\em hep-th/9709167}.

\bibitem{suzuki-twoa}
A.~T. Suzuki and A.~G.~M. Schmidt.
\newblock Solutions for a massless off-shell two loop three point vertex.
\newblock {\em hep-th/9712104}.

\bibitem{suzuki-twoa1}
A.~T. Suzuki and A.~G.~M. Schmidt.
\newblock An easy way to solve two-loop vertex integrals.
\newblock {\em Phys. Rev. D}, 58:047701, 1998.

\bibitem{suzuki-tensor}
A.~T. Suzuki and A.~G.~M. Schmidt.
\newblock Feynman integrals with tensorial structure in the negative
  dimensional integration scheme.
\newblock {\em Eur. Phys. J.}, C-10:357--362, 1999.

\bibitem{suzuki-three}
A.~T. Suzuki and A.~G.~M. Schmidt.
\newblock Negative dimensional approach for scalar two loop three-point and
  three-loop two-point integrals.
\newblock {\em Canad. Jour. Physics}, 78:769--777, 2000.

\bibitem{suzuki-massive}
A.~T. Suzuki and A.~G.~M. Schmidt.
\newblock Massless and massive one-loop three-point functions in negative
  dimensional approach.
\newblock {\em Eur. Phys. J.}, C-26:125--137, 2002.

\bibitem{vardi1}
I.~Vardi.
\newblock Integrals, an {I}ntroduction to {A}nalytic {N}umber {T}heory.
\newblock {\em Amer. {M}ath. {M}onthly}, 95:308--315, 1988.

\bibitem{zagier2}
D.~Zagier.
\newblock Hyperbolic manifolds and special values of {D}edekind
  {Z}eta-functions.
\newblock {\em Inv. {M}ath.}, 83:285--301, 1986.

\bibitem{zinn-1}
J.~Zinn-Justin.
\newblock {\em Quantum {F}ield {T}heory and {C}ritical {P}henomena}.
\newblock Clarendon Press, Oxford, 4th edition, 2002.

\end{thebibliography}
\end{document}